\documentclass[sn-mathphys,Numbered]{sn-jnl}

\usepackage{graphicx}%
\usepackage{adjustbox}
\usepackage[T1]{fontenc}
\usepackage[table]{xcolor}
\usepackage{setspace}
\usepackage{caption}
\usepackage[euler]{textgreek}
\usepackage{csquotes}

\usepackage{multirow}%
\usepackage{amsmath,amssymb,amsfonts}%
\usepackage{amsthm}%
\usepackage{cleveref}
\newtheorem{assumption}{Assumption}
\usepackage{mathrsfs}%
\usepackage[title]{appendix}%
\usepackage{xcolor}%
\usepackage{textcomp}%
\usepackage{manyfoot}%
\usepackage{booktabs}%
\usepackage{algorithm}%
\usepackage{algorithmicx}%
\usepackage{algpseudocode}%
\usepackage{listings}%

\theoremstyle{thmstyleone}%
\newcommand{\specialcell}[2][c]{%
  \begin{tabular}[#1]{@{}c@{}}#2\end{tabular}}
\theoremstyle{thmstyletwo}%

\theoremstyle{thmstylethree}%

\begin{document}

\title[An equilibrium-seeking search algorithm for integrating large-scale activity-based and dynamic traffic assignment models]{An equilibrium-seeking search algorithm for integrating large-scale activity-based and dynamic traffic assignment models}


\author*[1,2]{\fnm{Serio} \sur{Agriesti}}\email{serio.agriesti@aalto.fi}

\author[1]{\fnm{Claudio} \sur{Roncoli}}\email{claudio.roncoli@aalto.fi}
\equalcont{These authors contributed equally to this work.}

\author[3]{\fnm{Bat-hen} \sur{Nahmias-Biran}}\email{bathennb@ariel.ac.il}
\equalcont{These authors contributed equally to this work.}

\affil*[1]{\orgdiv{Department of Built Environment}, \orgname{Aalto University}, \orgaddress{\street{Otakaari 1}, \city{Espoo}, \postcode{02150}, \country{Finland}}}

\affil[2]{\orgdiv{FinEst Centre for Smart Cities}, \orgname{Tallinn University of Technology}, \orgaddress{\street{Ehitajate tee 5}, \city{Tallinn}, \postcode{19086}, \country{Estonia}}}

\affil[3]{\orgdiv{The Porter School of the Environment and Earth Sciences and the School of Social and Policy Studies}, \orgname{Tel Aviv University}, \orgaddress{\street{Ramat Aviv}, \city{Tel-Aviv}, \postcode{6997801}, \country{Israel}}}


\abstract{This paper proposes an iterative methodology to integrate large-scale behavioral activity-based models with dynamic traffic assignment models. The main novelty of the proposed approach is the decoupling of the two parts, allowing the ex-post integration of any existing model as long as certain assumptions are satisfied. A measure of error is defined to characterize a search space easily explorable within its boundaries. Within it, a joint distribution of the number of trips and travel times is identified as the equilibrium distribution, i.e., the distribution for which trip numbers and travel times are bound in the neighborhood of the equilibrium between supply and demand. 
The approach is tested on a medium-sized city of 400,000 inhabitants and the results suggest that the proposed iterative approach does perform well, reaching equilibrium between demand and supply in a limited number of iterations thanks to its perturbation techniques. Overall, 15 iterations are needed to reach values of the measure of error lower than 10\%. The equilibrium identified this way is then validated against baseline distributions to demonstrate the goodness of the results.}




\maketitle

\section{Introduction}\label{sec1}

Activity-based travel demand models and agent-based models (ABM) have some key advantages over more traditional four-step models, and their popularity is increasing accordingly. These advantages include framing disaggregate demand patterns, with individual choices based on socio-demographic features, and a traffic supply simulated down to the single agent \cite{ABMLit}. Large-scale traffic assignment (TA) models instead can be as detailed as the input demand, with mesoscopic and microscopic applications considering each vehicle as a single agent in its iterations with surrounding elements \cite{AimsunPaper, VissimPaper, SUMOPaper}. 
Nevertheless, the higher the considered disaggregation, the more these models become data-intensive \cite{SynthPop, ABMLit} and very difficult to calibrate \cite{Preprint}. Most of the activity-based models (all, to the authors' knowledge) require a synthetic population with enough data to characterize a population of agents, usually counting hundreds of thousands of entries \cite{SynthPop}, while the calibration process may involve hundreds of parameters \cite{Preprint}. This complexity is hindering the wider adoption of activity-based ABM models and, in general, disaggregate demand/supply modeling. This paper tries to tackle this complexity by decoupling the traffic assignment problem from the activity-based simulation and designing a methodology to integrate these two models ex-post, i.e., integrating existing TA with an existing activity-based model, with no need for recalibration of any of the underlying parameters. By existing, it is meant a model that has previously been built for other applications.

The problem of integrating disaggregated demand and supply models is not new in literature and has been tackled in different ways. \citep{2008Integration} conceptualize a framework to integrate a dynamic TA model with an activity-based one, exploiting iterations and assessing convergence through travel times \textbf{or} number of trips. This is, to the authors' knowledge, the first work formalizing this kind of integration but the stream of literature that followed did not result in a unified approach and has been limited to a handful of works, the majority of which are geographically located in the US. Two likely causes are the effort needed to build not one but two large-scale models and the constraints caused by the adopted tools (which may have an impact on the kind of designed solutions). More importantly, no work focuses on the problem of integrating existing models, with the additional constraint of having fixed model parameters. The research gap is still topical and scarcely addressed, as \cite{newlit} finds that the integration of demand generation and traffic assignment still needs a robust solution to be found. 

To properly formalize an integration and the methodology behind it, one of the key elements is the choice of the Measure of Error (MoE), through which to assess the goodness of the equilibrium between the demand and supply modules.
A stream of works adopts Behavioral User Equilibrium (BUE) as a measure of error, defined as the number of agents who have reason to change their behavior \cite{AgBM-DTALite, WashingtonDC}. In \cite{AgBM-DTALite}, the MoE is theorized and applied to reach equilibrium after a set of iterations; still, the MoE values rely on a dedicated behavioral adaptation experiment to be properly tuned and are therefore not easily transferable. In \cite{WashingtonDC}, a similar integration approach is adopted but it is not clear that the analyzed scenario (active traffic management) is further assessed through iterations once the BUE is applied.
\citep{Washington2}, instead, adopt as MoE the number of infeasible agents, i.e., the number of agents having one item in the daily activity schedule lasting less than a minimum duration (e.g., staying at the workplace for an unrealistic short time). These thresholds are defined through a mobility survey but may fail to frame disrupted or flexible scheduling due to new trends (e.g., hybrid working). Both MoEs may also struggle while assessing the integration performance in future scenarios that defy historical patterns (e.g., automated driving).

Not all the works on the subject focus on or detail the MoE. Another stream is dedicated instead to the data exchange between the two models, often happening mid-simulation and requiring dedicated interfaces. \citep{Phoenix}, for example, set up a data exchange between the two models happening every time a destination is reached, for the agents to update their behavior depending on the actual cost of the previous trip. \citep{Integration} categorize different integration types based on the nature of data exchange (in parallel or iterative) and the developed ad-hoc modules. The authors define five levels: sequential integration (L0), no real-time information (L1), only pre-trip information (L2), pre-trip and en route information with route diversion only (L3), and pre-trip and en route information with full activity-travel choice adjustments (L4). Already from L1, information between the two models is exchanged at regular intervals, e.g., every minute. The authors test a scenario falling in the L3 category. Other works addressing the integration problem are \cite{2018Heinrichs, SimAgent}.

The iterative approach followed in this paper falls in the L0 category \cite{Integration}, since one of its strengths is the applicability to any existing model falling either in the TA or in the activity-based realm, with no need for modifications. This, in turn, requires the absence of ad-hoc exchange information modules, to ensure the applicability to any model regardless of the simulation architecture. The trade-off sacrifices the communication to allow the integration between existing models, allowing, for example, to incorporate activity-based behavioral elements into large TA models relying on an aggregated demand. Another strength of the L0 approach is that it does not require specific software and applies to any meso/microscopic TA and activity-based tool. This, in turn, allows the usage of a multitude of TA tools, depending on the case study at hand (e.g., hybrid models with both meso- and microscopic degrees of precision). Besides, the defined MoE differs from others in the literature as it is not reliant on additional information and can be easily transferred and compared among case studies or applied to future scenarios.

This flexibility is the main strength when compared to existing integrated solutions. One example is MATSim \cite{MATSimZurich2,MATSimManual}, which, despite being capable of simulating both disaggregate demand and supply and enjoying multiple applications in Europe \cite{ViennaMatsim,CroatiaMatsim, BerlinMatsim}, has some key differences from the proposed approach. The main ones are the lower level of detail dedicated to the behavioral model of each individual in the considered population and the TA module being limited to mesoscopic applications. Besides, the reliance on historic data for the convergence of supply and demand lowers again the adaptability of said models to scenarios defying historical patterns (e.g., remote working, automated driving, etc.) \cite{MATSimManual}. 
On the other hand, purely modular models,i.e., representing one dimension of the activity-based versus TA dichotomy, able to run and produce results with no need for data exchange during each simulation, can be utilized within the proposed methodology, with no adaptation required (e.g., the activity-based CEMDAP \cite{CEMDAP}). 
Another example is SimMobility, which combines an activity-based module with a dedicated TA tool \cite{SimMobMT, TRR2, SingapSimMob}. Still, to the authors' knowledge, the SimMobility integration process, convergence, and analysis of the resulting equilibrium are not detailed in a scientific publication or elsewhere. This leaves a gap that limits replicability or transferability.
By presenting this methodological approach, this paper aims at providing researchers and practitioners with a feasible way to shift to activity-based modeling, by allowing the use of existing TA models regardless of the tool they are employing. This, in turn, should foster the shift towards ABM that is still lagging \cite{ABMLit}.  

The paper contributes to the current state-of-the-art by introducing a new iterative approach to tackle the integration of \textbf{existing} travel demand and traffic assignment models and by testing said approach for a large-scale scenario. It does so by defining an intuitive (i.e., physically meaningful) MoE and assessing its performance in framing the different dimensions of the modeling experiment. The MoE can be understood as the difference in traffic quantity across the network, and will be referred throughout the paper as $\Delta$Tf. A search space is characterized through it and explored to identify the most likely equilibrium distributions. Moreover, this paper extends previous work~\cite{mt-its2023}, in which we presented a preliminary version and results, by applying the proposed methodology to a dynamic traffic assignment and by improving the MoE (reducing its reliance on scale factors). This is a major improvement over \cite{mt-its2023}, where only static traffic assignment has been simulated. By applying the methodology to a large-scale case study, we validate the approach and provide the first, important, integration example within the European context. 

\begin{figure}[tb]
\centerline{\includegraphics[scale=0.20]{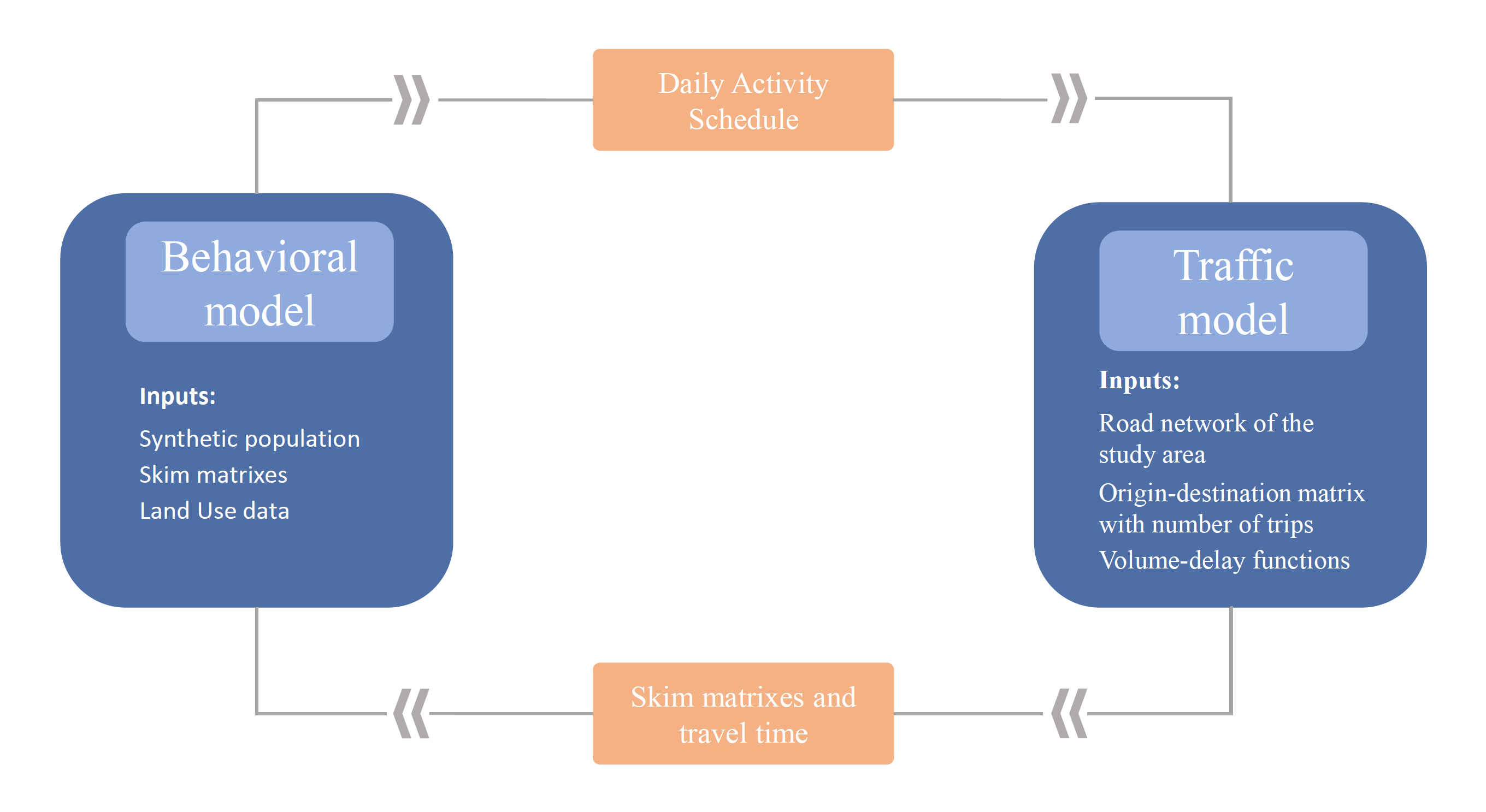}}
\captionsetup{belowskip=-10pt}
\caption{The proposed iterative approach - the gray thread represents the data exchange}
\label{Iterative Approach}
\end{figure}

\section{Methodology}\label{Methodology}

\subsection{Assumptions and theoretical background} \label{assumptions}
Before presenting the proposed algorithm, it is necessary to detail the underlying assumptions. 
The problem is to reach an equilibrium between demand and supply models across a set of iterations. Albeit both the models considered in this study entail a fair amount of complexity, the assumptions allowing us to hypothesize the existence of an equilibrium distribution for $\Delta$Tf are the following.

First, there is an inverse correlation between the number of trips and the travel time for each cell in the considered origin-destination OD matrix (i.e., an increase in travel time results in a decrease in the number of trips), which results in the following assumption.
\begin{assumption}\label{1F}
The inverse correlation between $n$ and $tt$ guarantees that there is an equilibrium point for each OD pair. This point is a combination of the calibrated parameters in the activity-based model and of the volume-delay functions (or equivalent) from the traffic assignment. This point lies along an unknown function f\textsubscript{x} within a defined area.
\end{assumption}

\begin{figure}[tb]
\centerline{\includegraphics[scale=0.15]{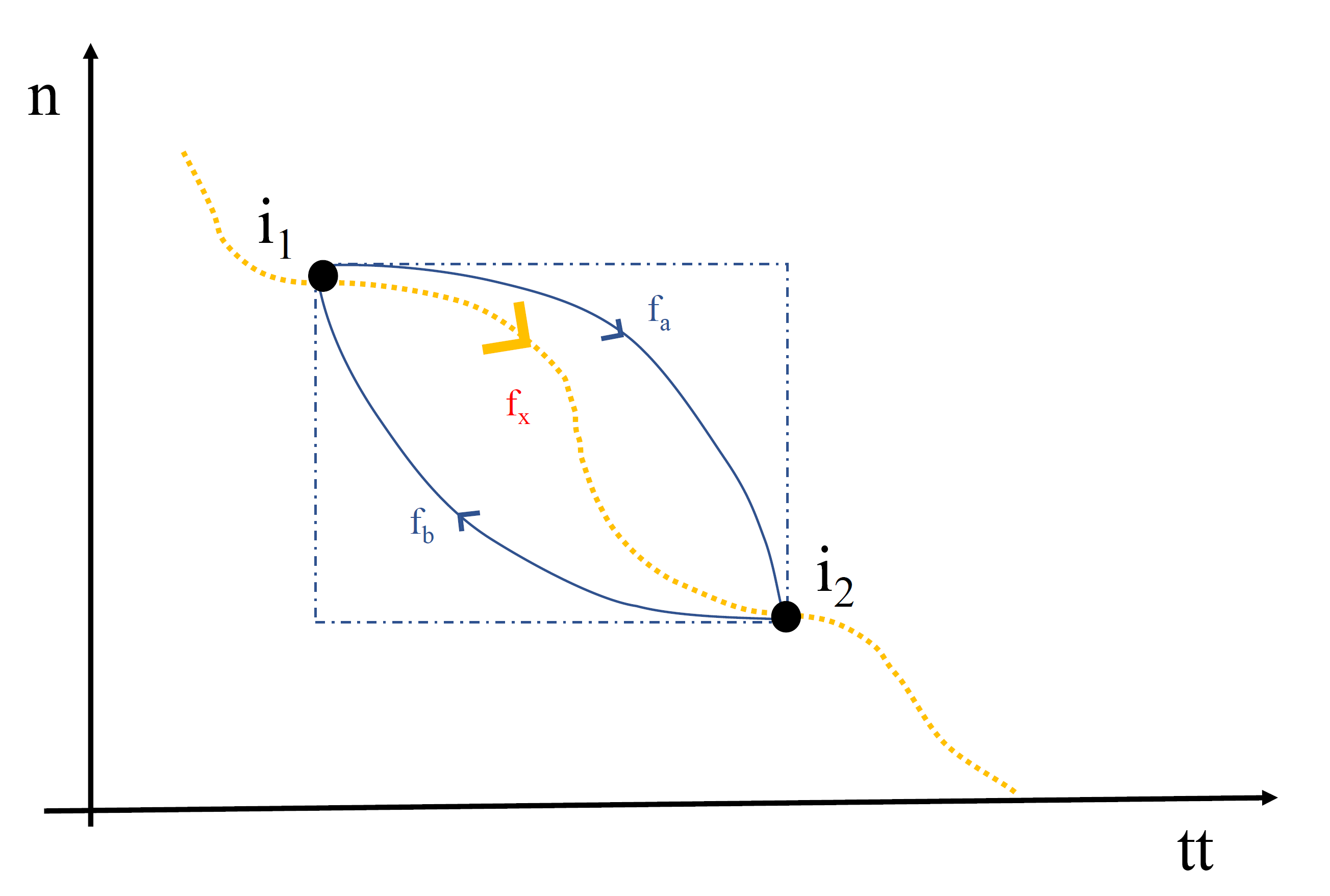}}
\captionsetup{belowskip=-10pt}
\caption{Assumption 1: Inverse correlation between travel time ($tt$) and number of trips ($n$) - For each OD pair, an increase in travel time (from iteration i\textsubscript{1} to iteration i\textsubscript{2} - through f\textsubscript{a,x}) results in a lower number of trips and vice-versa (i\textsubscript{2} to i\textsubscript{1} through f\textsubscript{b})}
\label{Ass1}
\end{figure}

\noindent Despite seemingly straightforward, this assumption is needed to ensure that our search space may not result in divergent behaviors and is fundamentally reflected in the utility functions ruling over the activity-based model and the typical functions of a TA algorithm. The relationship between travel time and the number of trips is not limited to certain behaviors (e.g., linear rather than quadratic), as long as the correlation is inverse. 
Assumption~\ref{1F} is sketched in Figure~\ref{Ass1}, where an unknown nonlinear monotonic function f\textsubscript{x} describes the relation between the travel times and the number of trips. Thanks to the assumption, the space in which this function may lie is bound by the functions f\textsubscript{1} and f\textsubscript{2}, which are also monotonic. In case no knowledge about the boundaries is available, f\textsubscript{1} and f\textsubscript{2} may take the shape of a square (the dotted lines in Figure~\ref{Ass1}). Points~i\textsubscript{1} and i\textsubscript{2} represent the two extreme points (e.g., minimum and maximum $n$ (i.e. number of trips) within a day, the latter for example being constrained by the number of agents).

\begin{figure}[tb]
\vspace{2mm}
\centerline{\includegraphics[scale=0.12]{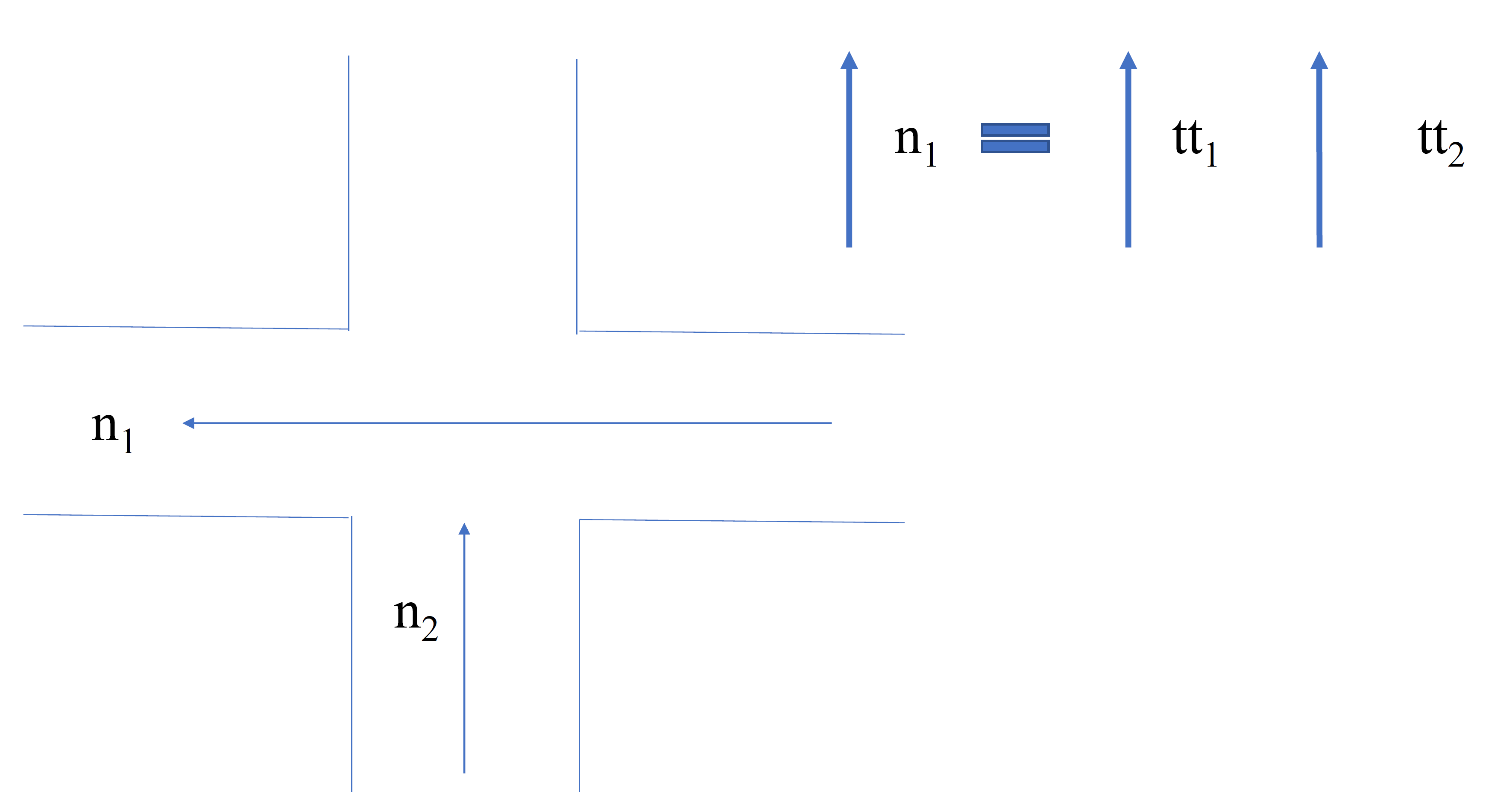}}
\vspace{2mm}
\captionsetup{belowskip=-10pt}
\caption{Network effects at a crossing}
\label{Network Effects}
\end{figure}

While the first assumption deals with the ($n$, $tt$) at the cell level in an OD matrix, the network effects and interactions across cells cannot be ignored. 
This is exemplified in Figure~\ref{Network Effects}, where an increase of $n$\textsubscript{1} will cause an increase of $tt$\textsubscript{2}, regardless of $n$\textsubscript{2}. Please note that this does not conflict with Assumption~\ref{1F}, because the first assumption only considers the relationship at the pair level (i.e., independent from boundary conditions).
The second assumption is then the following.
\begin{assumption}\label{2F}
The network effect on the equilibrium of each OD pair can be treated as a noise and approximated as an error term $\pm \Delta t$, which is not random but is unknowable a priori.
\end{assumption}
\noindent This error term is the result of a specific phenomenon, i.e., the conflicting traffic flows across one's path, as sketched in Figure~\ref{Network Effects}, and could be quantified for each OD pair only through a demanding number of experiments and iterations. Still, the assumption allows us to avoid such a task, whose feasibility is also limited by the stochasticity of the TA model.
Basically, the $\Delta t$ moves the equilibrium point along the existing function f\textsubscript{x}.
As a direct effect, we can formulate the last assumption.
\begin{assumption}\label{3F}
While the error term introduced in Assumption~\ref{2F} ($\Delta t$) leads to an intractable number of combinations of ($n$\textsubscript{i},$tt$\textsubscript{i}), the limited search space around f\textsubscript{x} guarantees that all the feasible equilibrium distributions will lie in the same neighborhood.
\end{assumption}
Starting at each OD pair level, the equilibrium between demand and supply from behavioral theory (law of demand \cite{transportEconomy}) guarantees the existence of an equilibrium point, for which the number of trips and travel times is balanced (even if said point correspond to a null number of trips), namely do not change until the system is perturbed.
Assumption~\ref{2F}, on the other hand, guarantees that the disruptions in travel time for each OD pair, caused by changes in the number of trips across the network, will only shift the equilibrium point for each OD pair, without affecting the inverse relationship, namely without changing the shape of the \textit{{n,tt}} function. This implies that an increase of trips in conflicting directions (Figure~\ref{Network Effects}) will shift the position of the equilibrium point on the f\textsubscript{x} function, but the equilibrium point will stay on the f\textsubscript{x} function (Figure~\ref{Ass1}). This, in turn, prevents diverging behaviors from emerging (i.e., a scenario for which, an increase in travel time may result in an increase in trips, that in turn would increase travel time and so on). 

To understand Assumption~\ref{3F} is to accept that stochastic large-scale models will not converge to a unique pair of matrixes for $n$ and $tt$ but that an equilibrium between demand and supply will be limited to a finite neighborhood space and will not have a diverging behavior. It is important to note that, since the presented work integrates two large-scale models for which no mathematical formulation is available, precisely framing the assumptions allows us to discuss the presented results later in the paper.

\subsection{Definition of a measure of error (MoE)}
\begin{figure}[tb]
\centerline{\includegraphics[scale=0.18]{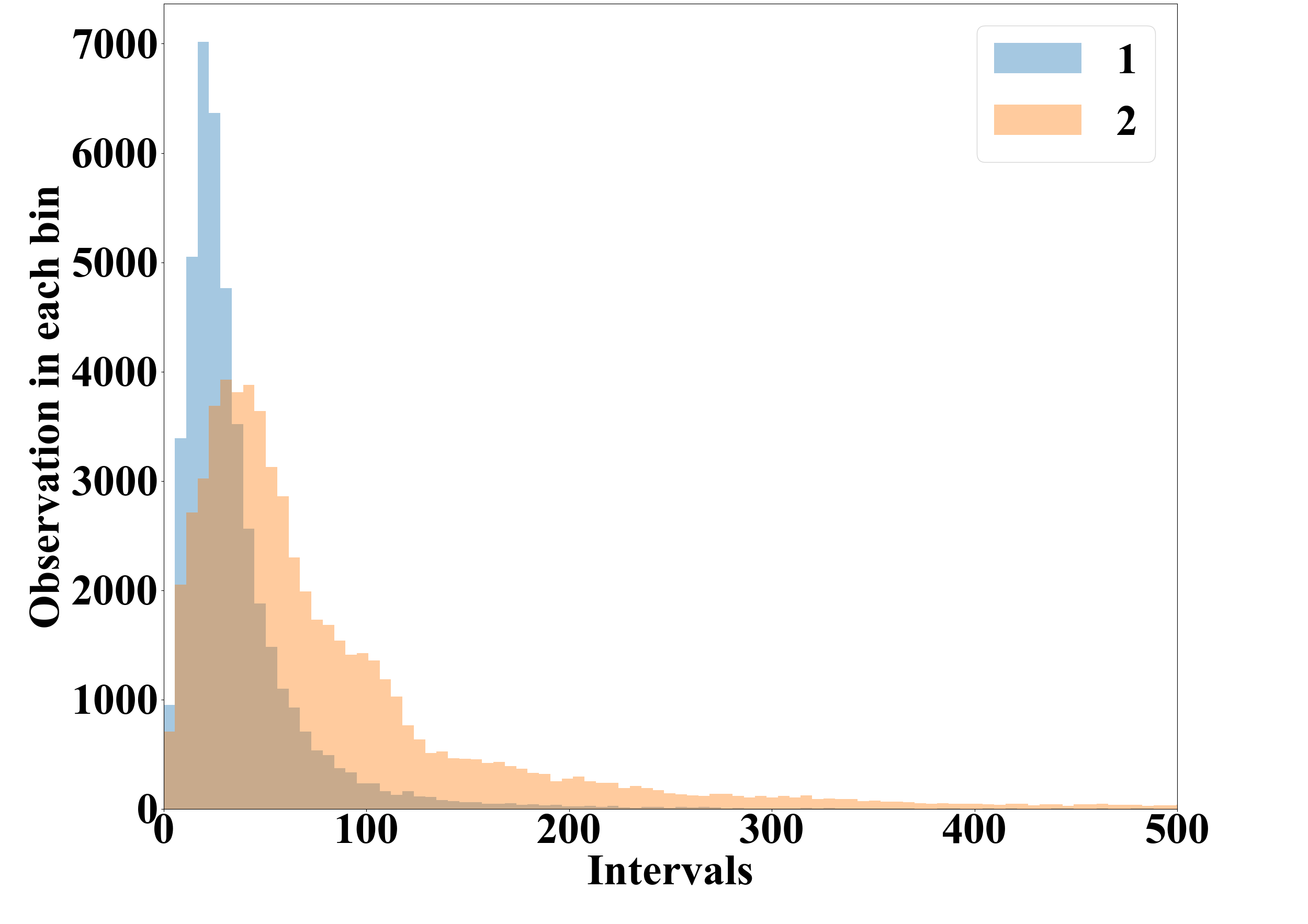}}
\captionsetup{belowskip=-10pt}
\caption{Area comparison between iteration 1 and 2}
\label{Single Iteration.png}
\end{figure}

In this work, we employ $n \cdot tt$ obtained from two consecutive iterations as the main metric that defines the MoE in terms of the difference in traffic patterns between consecutive iterations, which is denoted as~$\Delta$Tf. This is chosen for multiple reasons. First, it is not reliant on behavioral features of the case study at hand, being therefore easily transferable when compared, e.g., with the BUE \cite{AgBM-DTALite}. Furthermore, by not relying on behavioral information, it is also more easily applicable to assess future or disruption scenarios for which little information is available. This is also an advantage compared to the number of infeasible agents~\cite{Washington2}, which can hardly be defined for scenarios outside the available travel surveys. Finally, using $n \cdot tt$ allows framing the results of both models via $\Delta$Tf at once, i.e., employing $n$ from the activity-based model and $tt$ from the traffic assignment model, thus assessing each iteration in one single passage.
We define matrix $M_{od}$ as follows

\begin{equation}
\label{mod}
M^{O \times D} = N^{O \times D} \circ T^{O \times D},
\end{equation}

\noindent namely as the element-wise product
between the OD matrix $N$ and the travel time matrix $T$. 
Thus, each element $m_{od}$ of matrix $M$ is the resulting $n \cdot tt$ for each OD pair. 
It is worth highlighting again how the proposed methodology has been developed to ensure that any existing, calibrated TA or activity-based model may be integrated with their counterpart ex-post.  Being completely decoupled from any parameter involved in the calibration process of either model is then a highly desirable feature. $\Delta$Tf, being composed only of the outputs of each model (i.e., matrices $N$ and $T$), fits the scope perfectly. $\Delta$Tf is, therefore, the only one, to the authors' knowledge, that does not involve calibration parameters, is easily comparable across case studies, does not rely on historical data, and assesses both models at once. 
$\Delta$Tf is defined as: 

\begin{equation}
\label{eq:DeltaA}
    \Delta \textrm{Tf}^{~i} = \Delta A^{i} = |A^{i} - A^{i-1}| = \int |v^{i}(m_{od})-v^{i-1}(m_{od})| \;\; d(m_{od}) \quad \forall m_{od},
\end{equation}
\newline
\noindent where $A_{i}$ is the area under each distribution of values for an $M$ matrix, $i$ is the iteration index, $od$ represents each origin-destination pair for which the value $m_{od}$ is computed, and $v$ is the number of measurements (or cells) whose value falls in a certain range of $n \cdot tt$, as it will be explained in the following (refer also to Figure~\ref{Single Iteration.png}). Calculating the difference of areas (and of $v$) in absolute terms allows us to frame the overall \enquote{mismatch} between two matrices computed for successive iterations and avoid opposite differences to cancel each other out. 
To calculate $v$, we start by defining the interval size $u$ as

\begin{equation}
u = \frac{\textrm{max}(M)-\textrm{min}(M)}{L},
\end{equation}
\newline
\noindent where $L$ is the number of intervals; such intervals are equally distributed between the minimum and the maximum value of $M$, and are indexed by $l = 1...L$.
The number of measurements in each bin, $v(m_{od})$, is calculated via

\begin{equation}
\label{eq:v}
v^{l}(m_{od}) = \sum\limits_{1}^{O}\sum\limits_{1}^{D}\tau^l_{od}
\quad \forall l=1,\ldots,L,
\end{equation}
where

\begin{equation}
\tau^l_{od} =\begin{cases} 
   1, & \textrm{if } \, u \cdot (l-1) \leq m_{od} < l \cdot u \\
   0, & \textrm{otherwise}
\end{cases}
\quad \forall l=1,\ldots,L, o=1,\ldots,O, d=1,\ldots,D.
\end{equation}
\newline

\noindent Since $\Delta$Tf counts the values falling within each range, each OD pair has the same weight, and the mismatch is calculated only as the difference in trips $\cdot$ minutes (unit of measure of $m_{od}$ elements). $\Delta$Tf can be formulated as a percentage.

This approach is chosen over an approximation through kernel densities \cite{mt-its2023} to reduce the reliance of the results on arbitrary parameters such as the smoothing factor (which may actually change the shape of the curve and is therefore not just a scaling factor such as the bin range). To characterize the distribution through bins means to calculate how many cells in the $M$ matrix fall within a certain value interval (Equation~\ref{eq:v}). Comparing these matrixes between iterations allows us to assess the \enquote{traffic quantity} in trips $\cdot$ minutes difference. For example, the values $v$ for the first bin in Figure~\ref{Single Iteration.png} indicate how many cells in the $M$ matrix fall within 0-5 trips $\cdot$ minutes for iterations 1 and 2. The difference in terms of $v$ can then be quantified in trips $\cdot$ minutes as well.
$\Delta$Tf is calculated in relative terms instead, since cases with known distributions of $tt \cdot n$ at an urban scale are rare. The comparison is carried out on the areas from each pair of consecutive iterations.

\subsection{Characterization of the search space and equilibrium search}
\begin{figure}[tb]
\centerline{\includegraphics[scale=0.14]{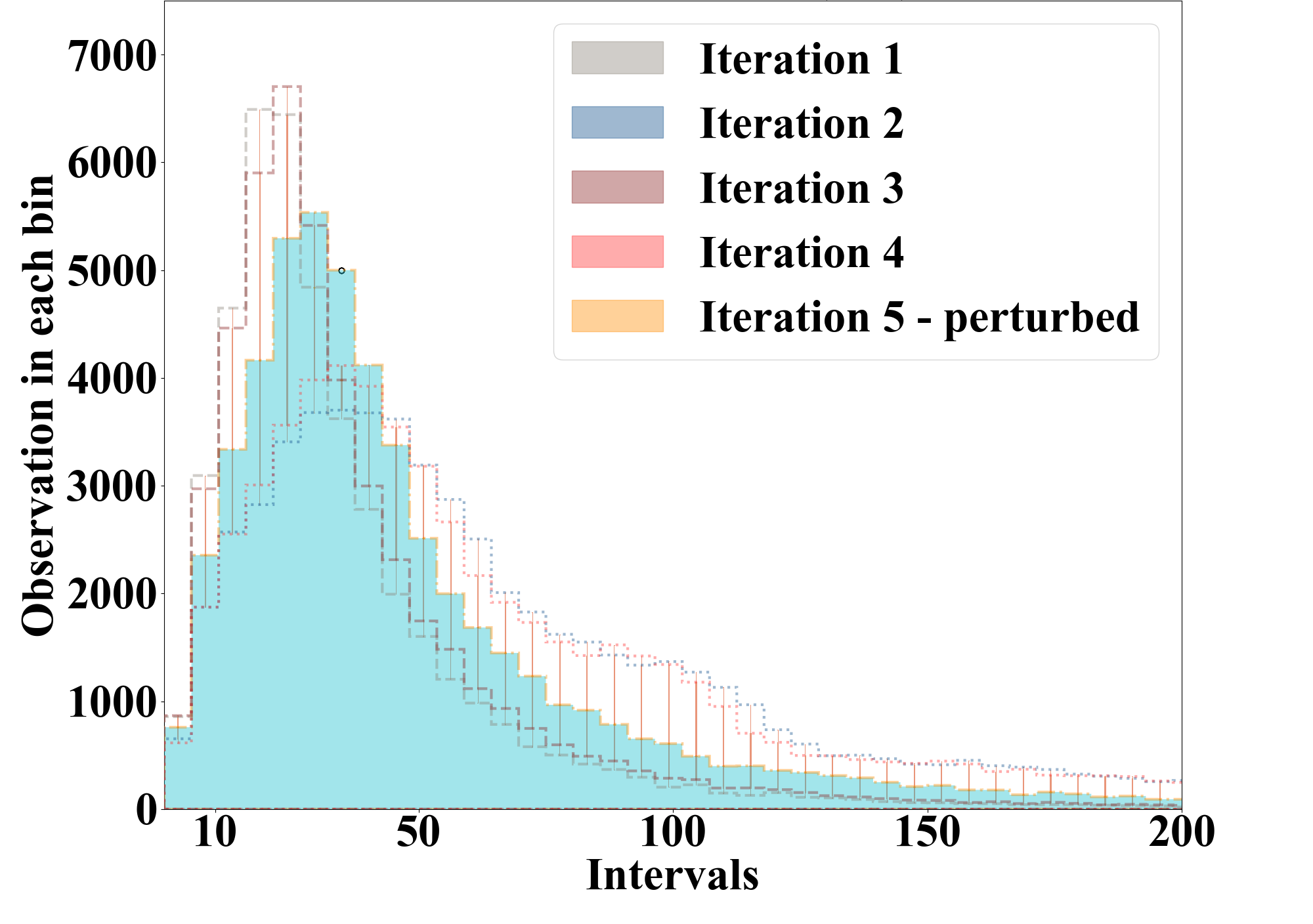}
\includegraphics[scale=0.14]{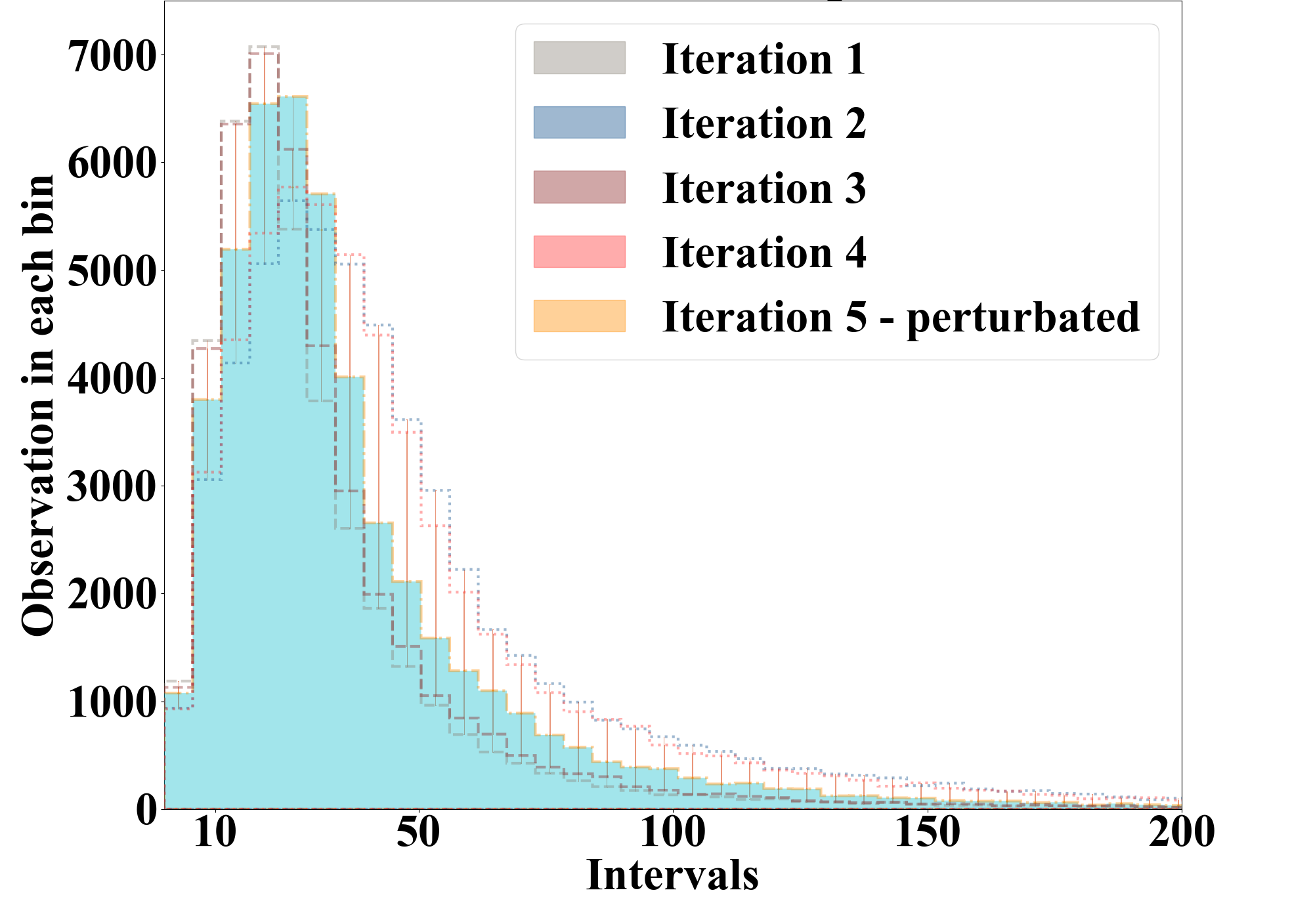}}
\captionsetup{belowskip=-10pt}
\caption{First round of iterations - morning (left) and afternoon peak (right); perturbed iterations shaded in light blue}
\label{Static Iterations}
\end{figure}

By simply running one iteration from virtually any starting point within the search space (Figure~\ref{Ass1}), a first estimation of the boundaries f\textsubscript{a} and f\textsubscript{b}, becomes available. In fact, based on the assumptions reported above, it is possible to state that an equilibrium distribution characterizing the equilibrium $n \cdot tt$ values for each OD cell lies within the boundaries set by the two histograms in Figure~\ref{Single Iteration.png}. A further set of iterations may be run to characterize the search space in a more precise way. Since no equilibrium is expected at this stage, additional iterations are only needed to validate the assumption and guarantee that no diverging behavior may arise.  
The first set of iterations is then exploited to explore the areas between the extremes within the search space (iterations 1-3 and 2-4 in Figure~\ref{Static Iterations}) and to identify the most promising $n \cdot tt$ distribution (iteration 5 in Figure~\ref{Static Iterations}). An exploring approach based on quantiles\footnote{Quantiles are cut points dividing the available observations into subsets with equal probabilities. In this work, any quantile lying between two observations is determined through linear interpolation} of the skim matrixes has been presented in \cite{mt-its2023}, where different time distributions have been tested and the most promising has been used as starting point for the further iterations.
To perturb travel times, i.e., to explore the search space by testing different distributions of $tt$ has been deemed the best approach rather than perturbing through different OD matrixes. In fact, to discretize $tt$ values for each cell in the matrix is both simpler (float values make sense) and more intuitive. 
A quantile sectioning of the search space is reported in more detail in Section~\ref{Case Study}. Since the perturbation is applied to the travel times, the quantile distribution is calculated over the skim matrixes from the first set of iterations (Figure~\ref{Static Iterations}).

Once the most promising $tt$ distribution is identified, the iterations are perturbed by injecting said distribution into the cycle. Starting from the new skim matrix $T$, a new set of iterations is carried out to assess the actual performance of the new distributions. The performance is assessed through $\Delta$Tf reported in Equation~\ref{eq:DeltaA}. As it will be shown in Section~\ref{Case Study}, the new $n \cdot tt$ distribution is expected to boast a smaller $\Delta$Tf and smaller variations between iterations, being carried out closer to the equilibrium. Should this not be the case, the search space should be further explored, with the aim of minimizing $\Delta$Tf and identifying the corresponding $tt$ distribution. Once a satisfying value of $\Delta$Tf is reached and no divergent behavior arises, the obtained results are validated against baseline distributions.

\subsection{A heuristic algorithm solution to a local search problem}
The algorithm at the core solves an optimization problem for which the cost to be minimized, $C$, is the value of $\Delta$Tf, namely the difference in the quantity of traffic calculated across iterations. This difference reflects how much each cell in the OD matrix oscillates around the point of equilibrium between demand and supply. To minimize this quantity means to minimize the cumulative distance across the matrix from the set of equilibrium points.

\begin{figure}[tb]
\centerline{\includegraphics[width=1.2\textwidth]{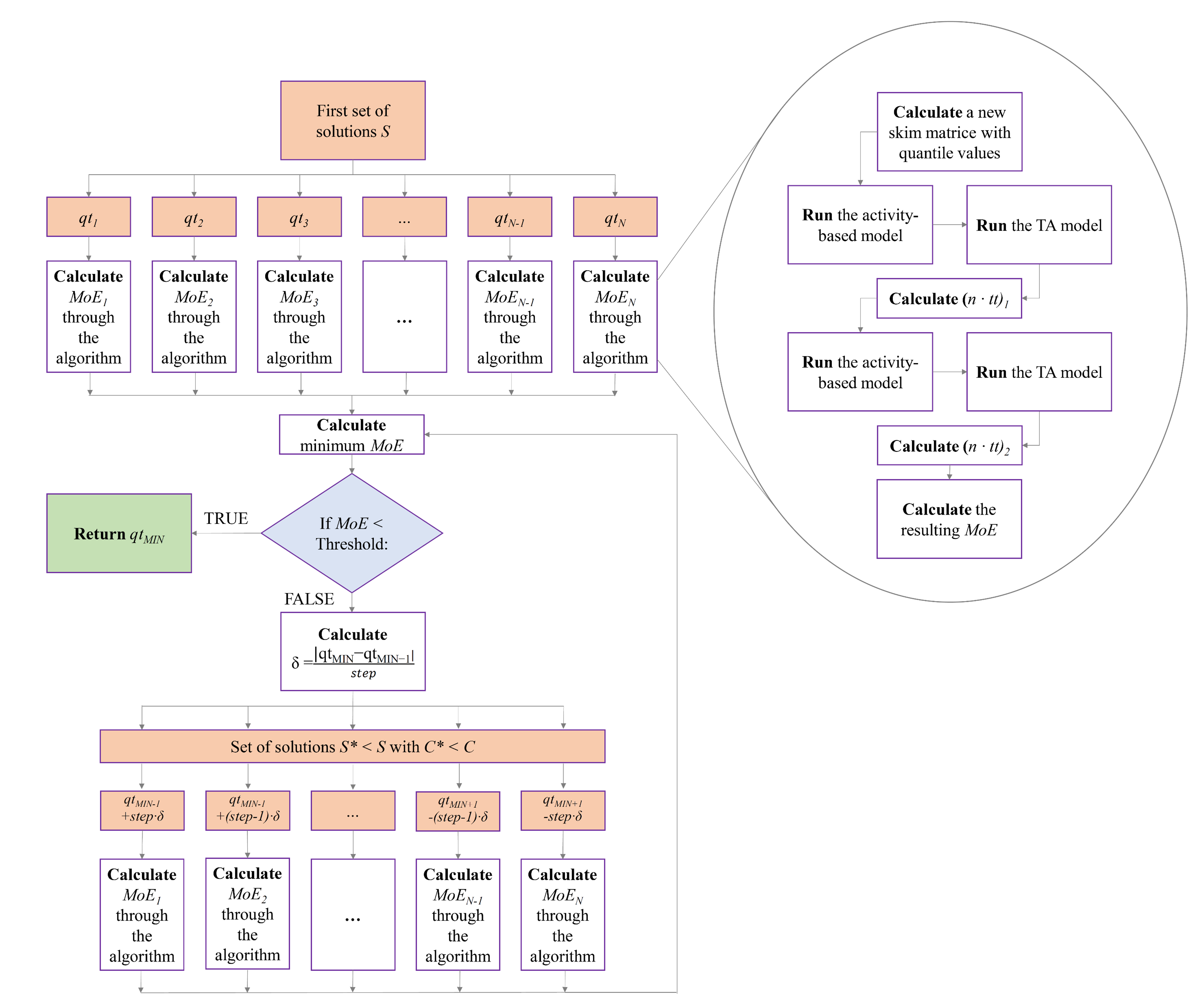}}
\captionsetup{belowskip=-8pt}
\caption{The proposed local search algorithm}
\label{algorithmqt.png}
\end{figure}

Figure~\ref{algorithmqt.png} summarizes the described heuristic iterative local search algorithm \cite{ILU, localSearch}. It is worth formalizing it also in general terms, to maximize the generality of the proposed approach and to properly define the theory behind it. 
Local search algorithms look for improved solutions $S^*$ through perturbations of the current solution $S$. They do so by scanning the neighborhood of each solution and scoring, through minimization of the cost, each new area \cite{localSearch}. It is worth noting that the final solution is a local minimum of $C$. In the presented case, the cost is the MoE, the neighborhood is characterized by adjacent distributions of $n \cdot tt$ and the perturbations are applied through quantiles of $tt$. By adopting a threshold value to terminate the iterations, we define an acceptance criterion and thus strongly favor intensification rather than diversification \cite{ILU}. We do so by exploiting the three initial assumptions, which allow us to state that \enquote{all the feasible equilibrium distributions will lie in the same neighborhood} (Assumption~\ref{3F}). Thus, diversification, i.e., the scanning of the search space for different local solutions, is less important than intensification, i.e., acceptance of only improved sets of solutions, namely with a lower $\Delta$Tf. This greatly increases the efficiency of the algorithm, allowing a parallel search streamlined to the lowest $\Delta$Tf available. To accept worse values of $\Delta$Tf through the search would instead disrupt the parallelization and greatly increase the number of branches and thus iterations needed (Figure~\ref{algorithmqt.png}). This concept is important as it justifies the approach and the final solution provided by the algorithm. 
According to Assumption~\ref{3F}, but also intuitively, the number of combinations of $n$ and $tt$ across a large OD matrix for which $C^* \rightarrow \min(C)$ is unfeasible to treat but also of scarce interest. It is enough to imagine how little increases of 1 minute in a single cell may impact the whole network and/or population of agents. And then project such a small increase to all the possible combinations of both $n$ and $tt$. It is impossible to escape such a conundrum due to the complexity of mathematical formulations in both activity-based and TA models on a large scale. So, the search for the exact distribution $n \cdot tt$ for which $C^* = \min(C)$ is not the objective of the algorithm (or rather, it is the general goal but the theoretical assumptions guarantee the goodness of the suboptimal solutions). By characterizing the problem as the search of a set of equilibrium points, each in a convex supply-demand space, and by then identifying the local minimum in the solution's neighborhood, the algorithm guarantees a solution in a limited amount of iterations, while justifying the search for a local rather than the global solution.

\section{Case Study}\label{Case Study} 
The case study is focused on the city of Tallinn, the capital of Estonia. The city is home to~$\sim$400,000 inhabitants and sees more than 1,100,000 trips in a typical day, of which almost half are carried out by motorized private transport. 

\subsection{The activity-based model}
\begin{figure}[tb]
\centerline{\includegraphics[width=0.8\textwidth]{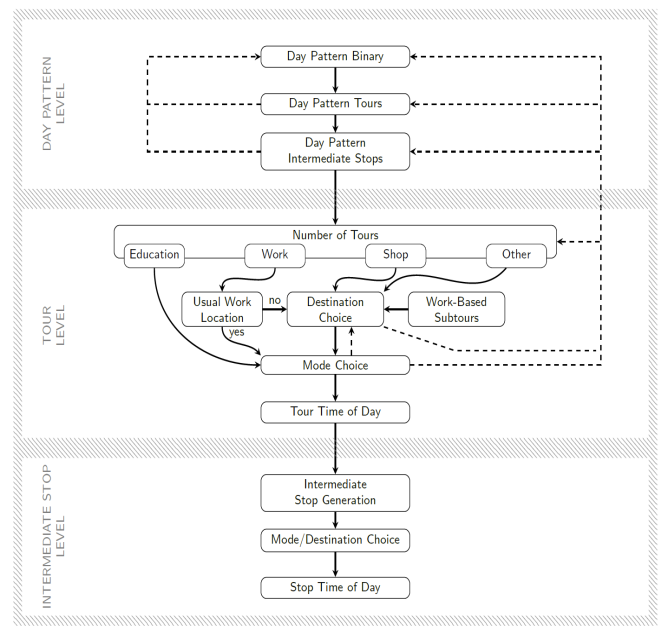}}
\captionsetup{belowskip=-8pt}
\caption{The nested logit structure used in the SimMobility-MTmodel; each level of the tree is characterized by a set of utility functions}
\label{nested logit bh.png}
\end{figure}

The considered behavioral activity-based model is built in SimMobility-MT \cite{SimMobMT} and exploits a series of nested-logit models to characterize the schedule of each one of the inhabitants (agents) down to the single choice within the day. This is performed by computing and comparing utilities at each level of the mobility tree (Figure~\ref{nested logit bh.png}) while exploiting the logsum concept \cite{Logsum} to tie the utilities at the top of the tree with the ones on lower branches. SimMobility-MT takes into account stochasticity through random seeds.

The model is built by defining 609 zones, which implies that 370,881 cells compose the $n \cdot tt$ matrix (as many as the OD pairs). The model has been built to reproduce relevant mobility patterns and socioeconomic features of the Tallinn population \cite{Preprint, SynthPop}. This, in turn, allows us to validate the equilibrium distribution against known baseline distributions (e.g., the aggregate OD matrixes, trip distributions, modes, etc.). It is important to highlight that no calibrated parameter will be modified during the integration, since the main aim of the methodological approach is to guarantee that existing models may be coupled together ex-post. 
\subsection{The traffic assignment model}
\begin{figure}[tb]
\centerline{\includegraphics[scale=0.20]{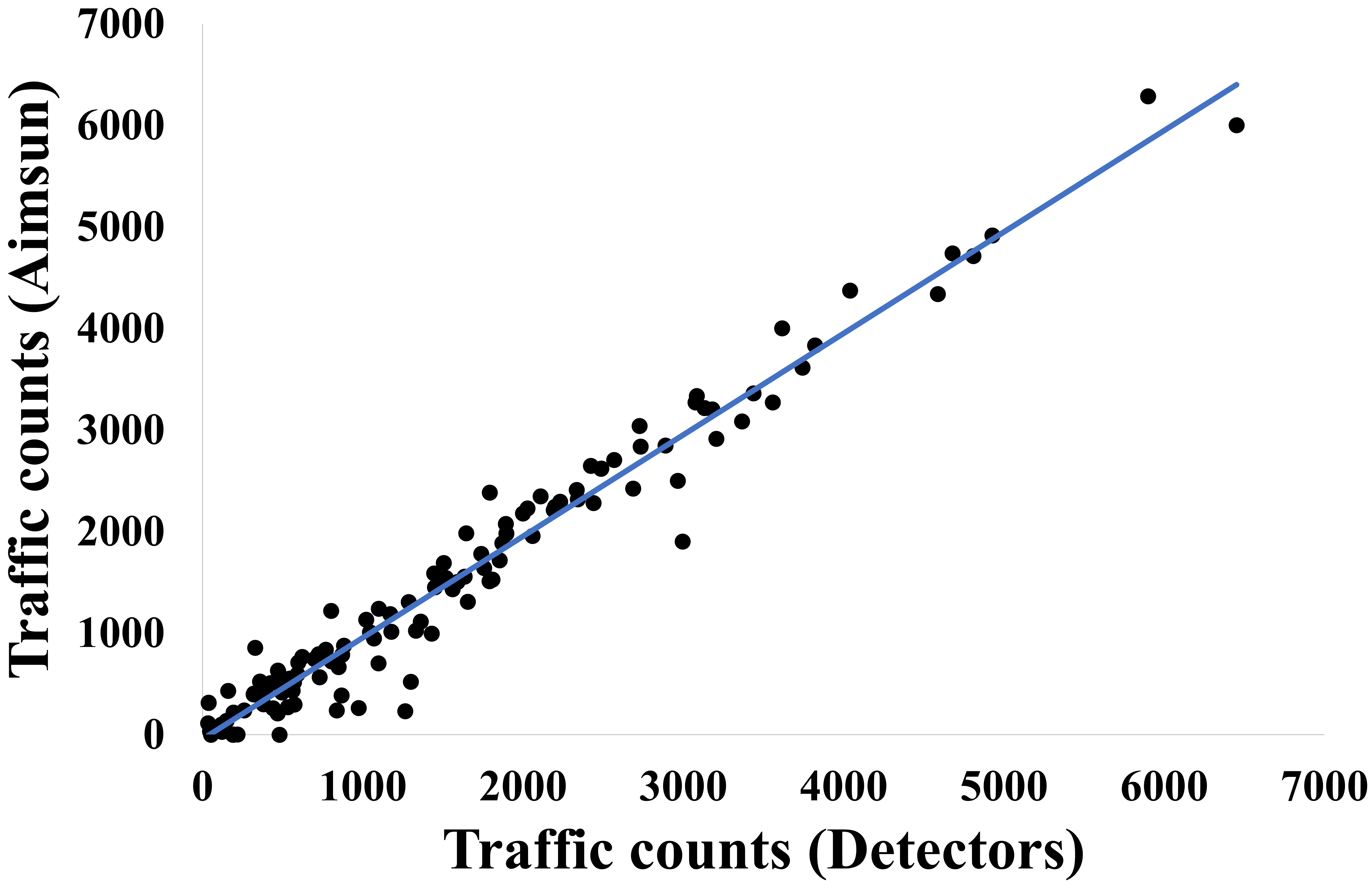}}
\captionsetup{belowskip=-10pt}
\caption{Match between simulation and baseline vehicle counts - Aimsun}
\label{Aimsun Rsq.png}
\end{figure}

A TA model is instead built with Aimsun~\cite{AimsunManual}. This model has the same set of origins-destinations as the one in SimMobility-MT and has been calibrated against traffic counts to match current guidelines \cite{LondonGuidelines} (Figure~\ref{Aimsun Rsq.png}). The calibrated values rule over different aspects of the traffic assignment, with functions such as volume-delay ones or turn-delay ones, defining the performance of each route in terms of travel time and as a function of volumes and capacities. Defining these elements and the desired speed distribution means calibrating the main dimensions of the macroscopic assignment problem. Items such as reaction time, C-Logit model parameters, and node connections rule instead over the behavior of single vehicles in the mesoscopic assignment.

\section{Numerical results}
\subsection{First set of iterations: Search space characterization}
\begin{figure}[tb]
\centerline{\includegraphics[scale=0.20]{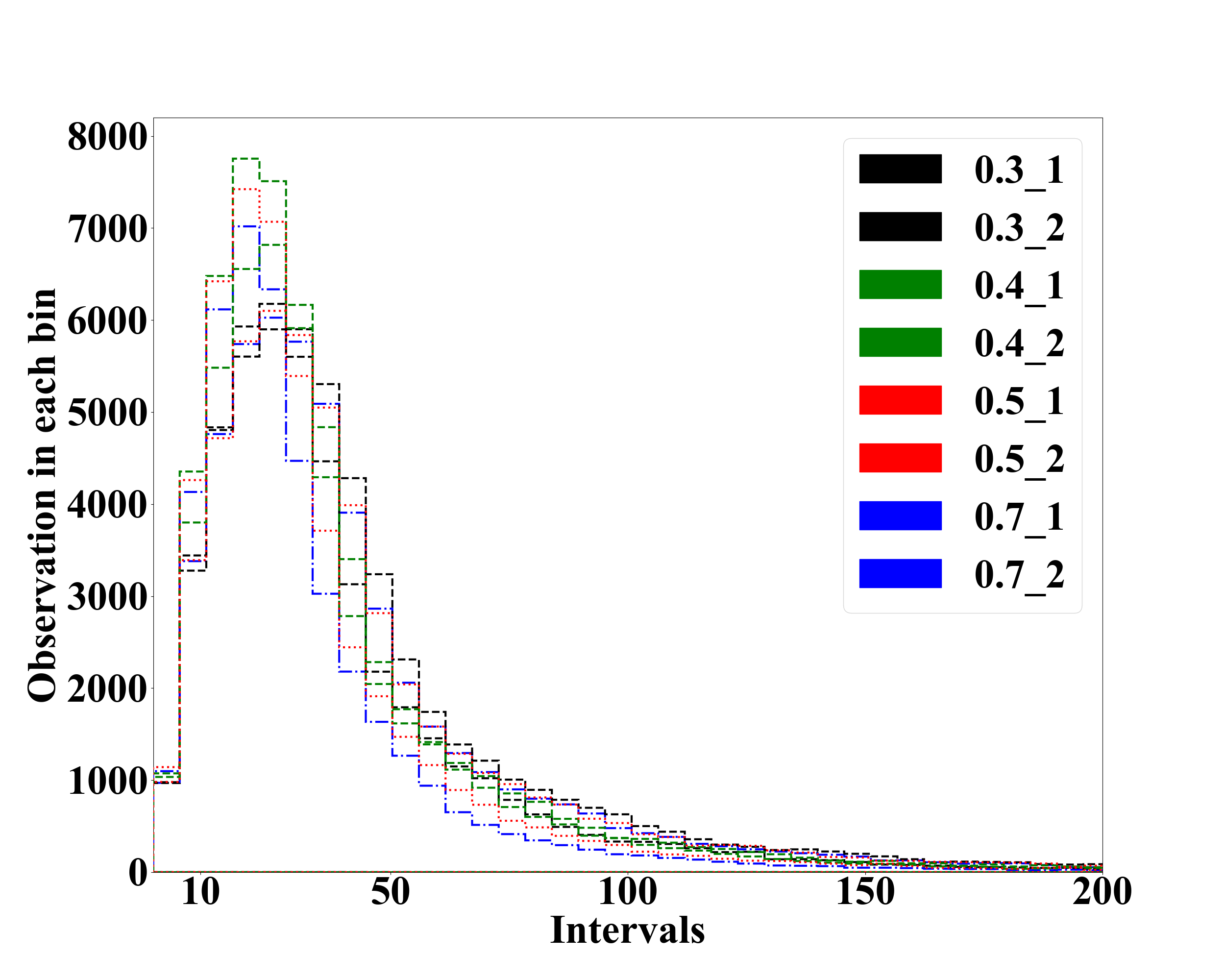}}
\captionsetup{belowskip=-10pt}
\caption{Sectioning of the search space through quantiles. On the x-axis, the first 200 of the 5000 equal intervals (i.e., bins of size $u$) between $\textrm{min}(n \cdot tt)$ and $\textrm{max}(n \cdot tt)$ are reported; on the y-axis, the number of measurements in the 378001 $n \cdot tt$ matrix falling within each interval is reported. For example, the number of values within the interval 0-5 trips $\cdot$ minutes corresponds to the first bin~(1000 on the y-axis).}
\label{Quantiles.png}
\end{figure}

The two models are run iteratively, starting from an estimate of travel times based on distances as input for the first iteration. The first four iterations produce the results shown in Figure~\ref{Static Iterations}, while the $\Delta$Tf values are reported in Table~\ref{MoE trend - 1}. 
These values represent the absolute difference in area between histograms calculated via Equation~\ref{eq:DeltaA}. From this first set of iterations, the search space is partitioned through quantiles as in Figure~\ref{Quantiles.png}. 

\begin{table}[tb]
\caption{$\Delta$Tf values across the first set of iterations}
\large
\begin{tabular}{m{2.5cm}m{2.5cm}}
\toprule
\textbf{Iterations} & \textbf{{$\boldsymbol{\Delta}$\textrm{Tf}}} \\
\midrule
1-2 & 0.71 \\
2-3 & 0.63 \\
3-4 & 0.6 \\
\botrule
\end{tabular}
\label{MoE trend - 1}
\vspace{-3mm}

\end{table}

\begin{figure}[tb]
\centerline{\includegraphics[scale=0.17]{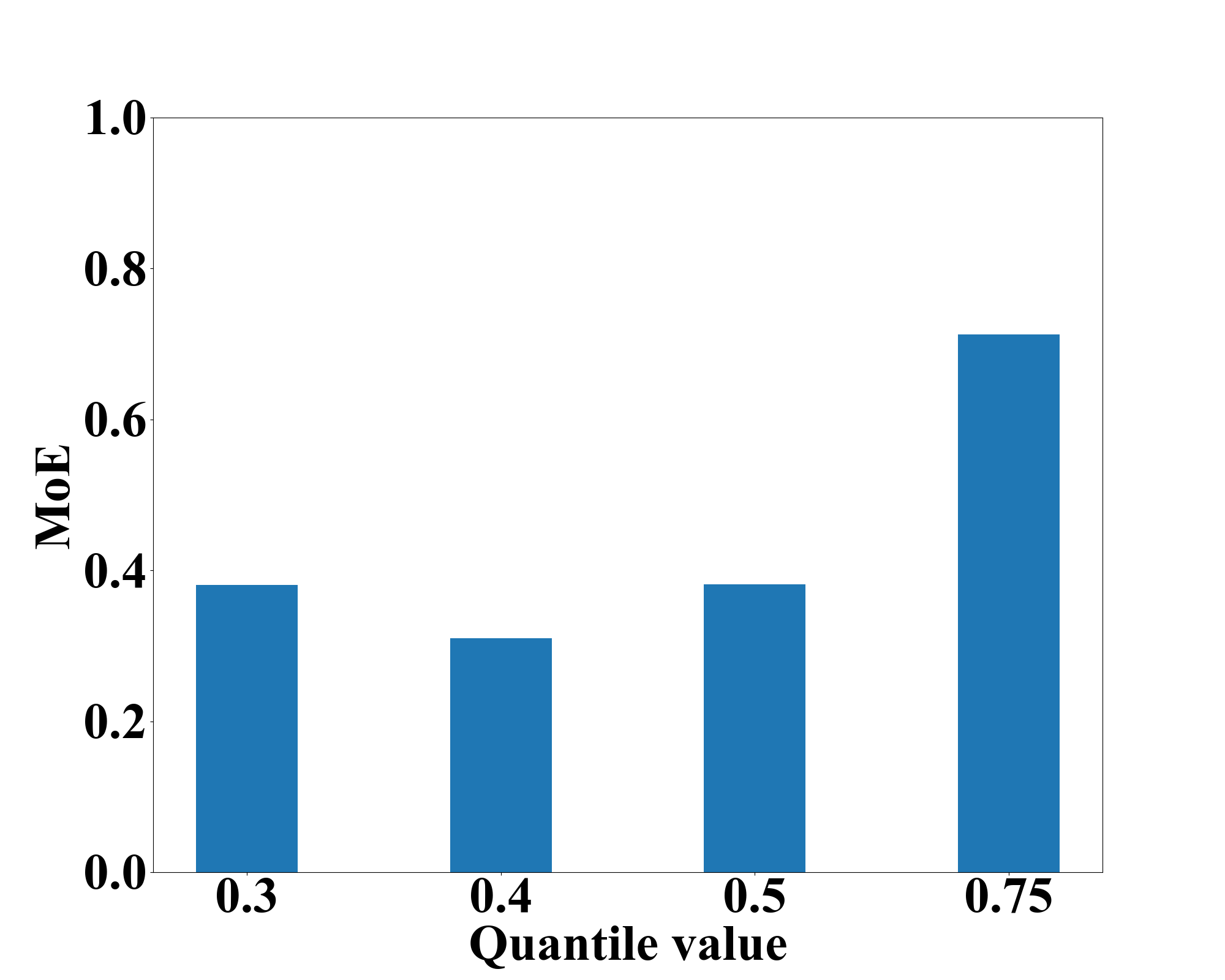}}
\captionsetup{belowskip=-10pt}
\caption{Sum of the $\Delta$Tf values calculated for the am and pm scenarios for each quantile value - Initial values}
\label{DA_sum_4}
\end{figure}

As mentioned in the methodology (Equation~\ref{eq:v}), intervals were defined within the interval~$\left[\textrm{min}(n \cdot tt),\textrm{max}(n \cdot tt)\right]$. A value of 5000 intervals was chosen so that each interval would be equal to around 5 trips $\cdot$ minutes. 5000 was defined by assessing the gains in histograms' definition (Figure~\ref{Static Iterations}) against computational times related to smaller values of $u$. 
For each quantile, the steps detailed in Figure~\ref{algorithmqt.png} are followed. Each quantile value is tested and the resulting $\Delta$Tf is evaluated. $\Delta$Tf is calculated through a linear function identifying a specific interval between two of the starting skim matrixes' distributions and then interpolating the value of $tt$ for each one of the 370881 cells \cite{quantile1, quantile2}. The result is another skim matrix that is then fed to the activity-based model for one cycle of iterations. If the quantile values do not result in a value of $\Delta$Tf lower than the acceptable threshold, the minimum $\Delta$Tf is used to identify the best-performing quantile, while its surroundings are further explored by repeating the cycle. A first set of quantile values equal to [0.3,0.4,0.5,0.75] has been tested and the resulting pattern indicated an overall smaller value of $\Delta$Tf around 0.4 when both morning and afternoon are considered (as reported in Figure~\ref{DA_sum_4}, where the $\Delta$Tf values for the morning and afternoon peak are summed).

A further exploration between 0.3 and 0.5 led to the partitioning of this space in intervals with a step of 0.03. In the end, $qt = 0.335$ was adopted as the final quantile value and starting $tt$ distribution for the next step of iterations (as described in \cite{mt-its2023}). The 0.335 quantile distribution is corresponding to the one shaded in light blue in Figure~\ref{Static Iterations}.
Simulating only identical pairs of values for the quantile from am and pm peaks (e.g. 0.3-0.3, 0.4-0.4) reduces the number of combinations to be tested. So, while it is possible to further tune the search through mixed combinations (e.g., quantiles of 0.3 for am and 0.4 for the pm peak scenario), minimizing the sum produces stable results with $\Delta$Tf lower than 10\%, as it will be shown.

\subsection{Second set of iterations: Perturbation}
\begin{figure}[tb]
\centerline{\includegraphics[scale=0.15]{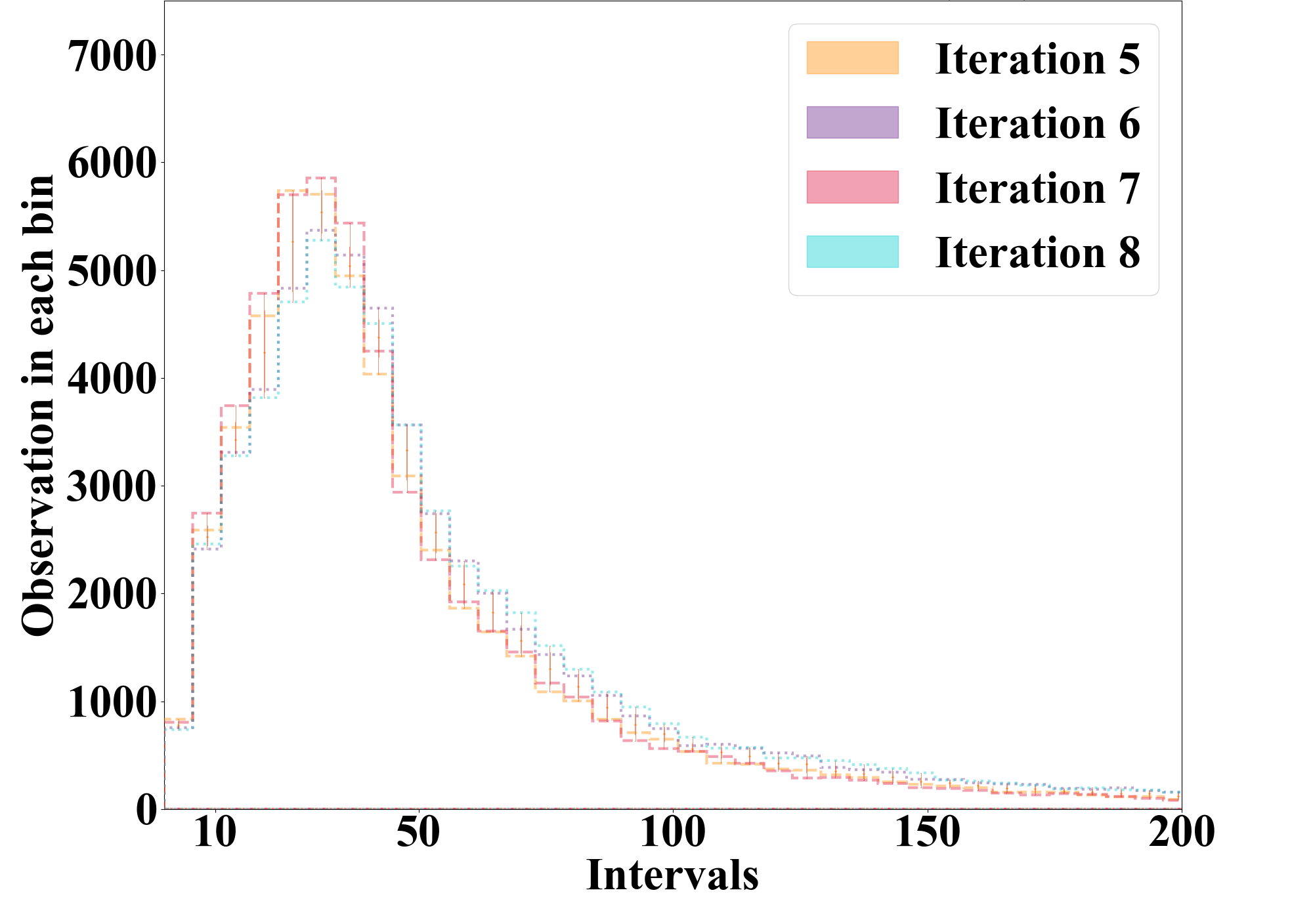}
\includegraphics[scale=0.15]{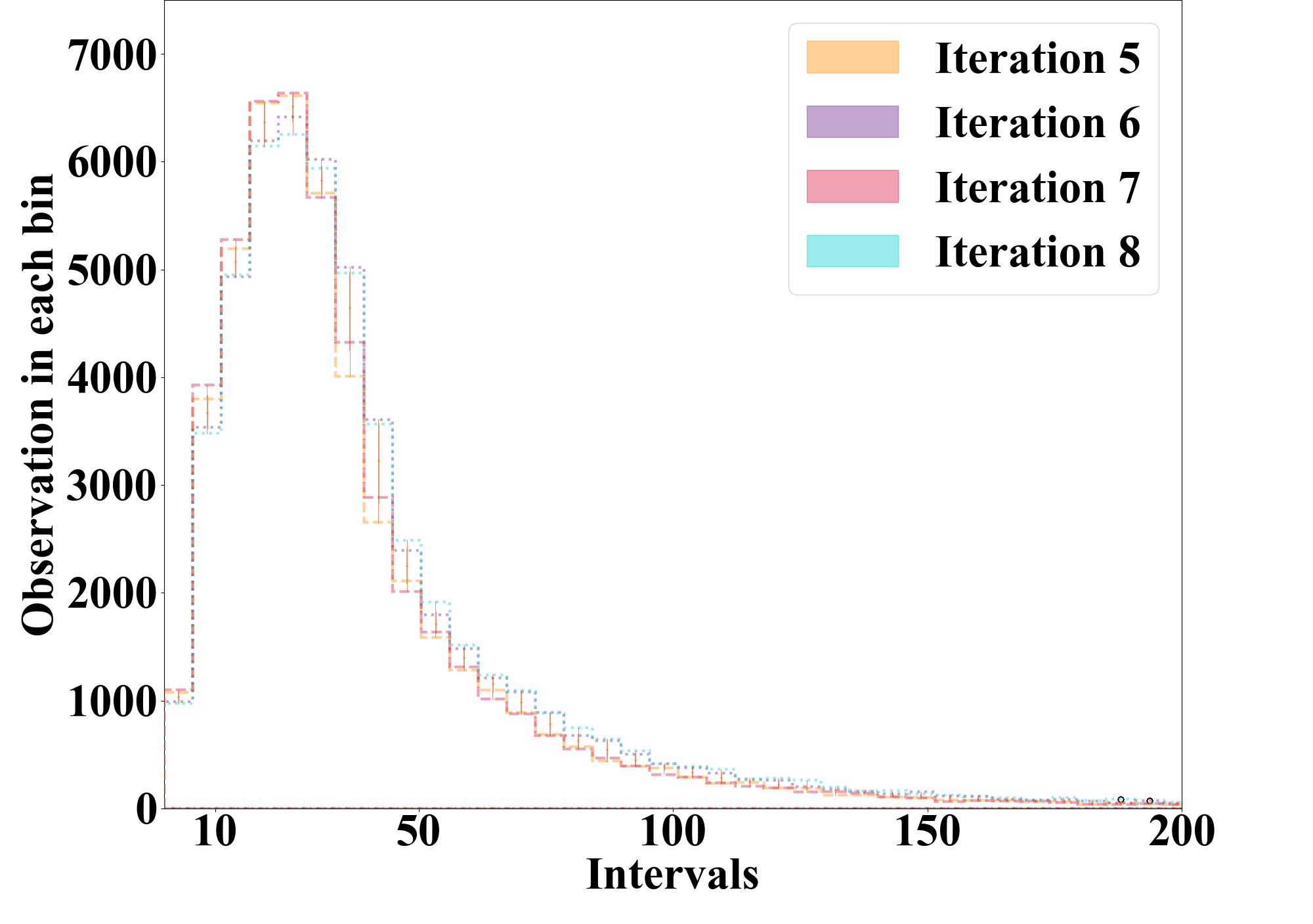}}
\captionsetup{belowskip=-10pt}
\caption{Static iterations after perturbation - morning (left) and afternoon peak (right)}
\label{Perturbed static.png}
\end{figure}

As shown in Figure~\ref{Static Iterations}, the identified travel time distribution is perturbed to match the 0.335 quantile and the resulting $n \cdot tt$ distribution indeed falls in a promising area of the search space. The resulting set of $n$ does not disrupt the equilibrium (e.g., by setting the iterations back to one of the two extremes as in Figure~\ref{Single Iteration.png}). To assess the stability of the equilibrium, four further iterations are carried out and the resulting $n \cdot tt$ distributions are reported in Figure~\ref{Perturbed static.png}. At this stage, a macroscopic assignment, if available can be used rather than a dynamic traffic assignment, to reduce computational time.

As can be noticed, the oscillations between iterations are sensibly smaller than in the previous set of iterations. This reflects the goodness of the perturbated skim matrix acting as the starting point for these new iterations. Besides, this validates the three assumptions presented in Section~\ref{Methodology}. The reduced oscillation around an equilibrium distribution of $n \cdot tt$ is reflected also in the value of $\Delta$Tf that drops to the values reported in Table~\ref{MoE trend - 2}. It should be highlighted again how these values represent the difference in percentages between areas.

\begin{table}[tb]
\caption{$\Delta$Tf values across the perturbed set of iterations}
\large
\begin{tabular}{m{2.5cm}m{2.5cm}}
\toprule
\textbf{Iterations} & \textbf{{$\boldsymbol{\Delta}$\textrm{Tf}}} \\
\midrule
4-5 & 0.34 \\
5-6 & 0.14 \\
6-7 & 0.15 \\
7-8 & 0.16 \\
\botrule
\end{tabular}
\label{MoE trend - 2}
\vspace{-3mm}
\end{table}

\subsection{Stability of the equilibrium}
As mentioned in the introduction, this work expands previous results by including a dynamic element in the TA. While the results in Table~\ref{MoE trend - 2} are based on a static TA (STA), without loss of generality it is possible to include a dynamic TA (DTA) in the simulation cycle to assess if the found equilibrium is impacted, and how, by it. To do so, the resulting skim matrix obtained via STA from iteration 8 becomes the first in a cycle of iterations utilizing DTA. The underlying hypothesis is based on the main difference between STA and DTA being the propagation of the congestion fronts \cite{ AimsunPaper,AimsunManual}, which in an urban environment tends to be localized and have ripple effects on the rest of the network. To do so, an additional set of 8 iterations is run and $\Delta$Tf is monitored. In case of diverging behavior, these iterations would be used to apply the algorithm and search for another perturbed distribution. It should be stressed that the methodology does not assume in any part the kind of assignment to be carried out.

\begin{figure}[tb]
\centerline{\includegraphics[scale=0.15]{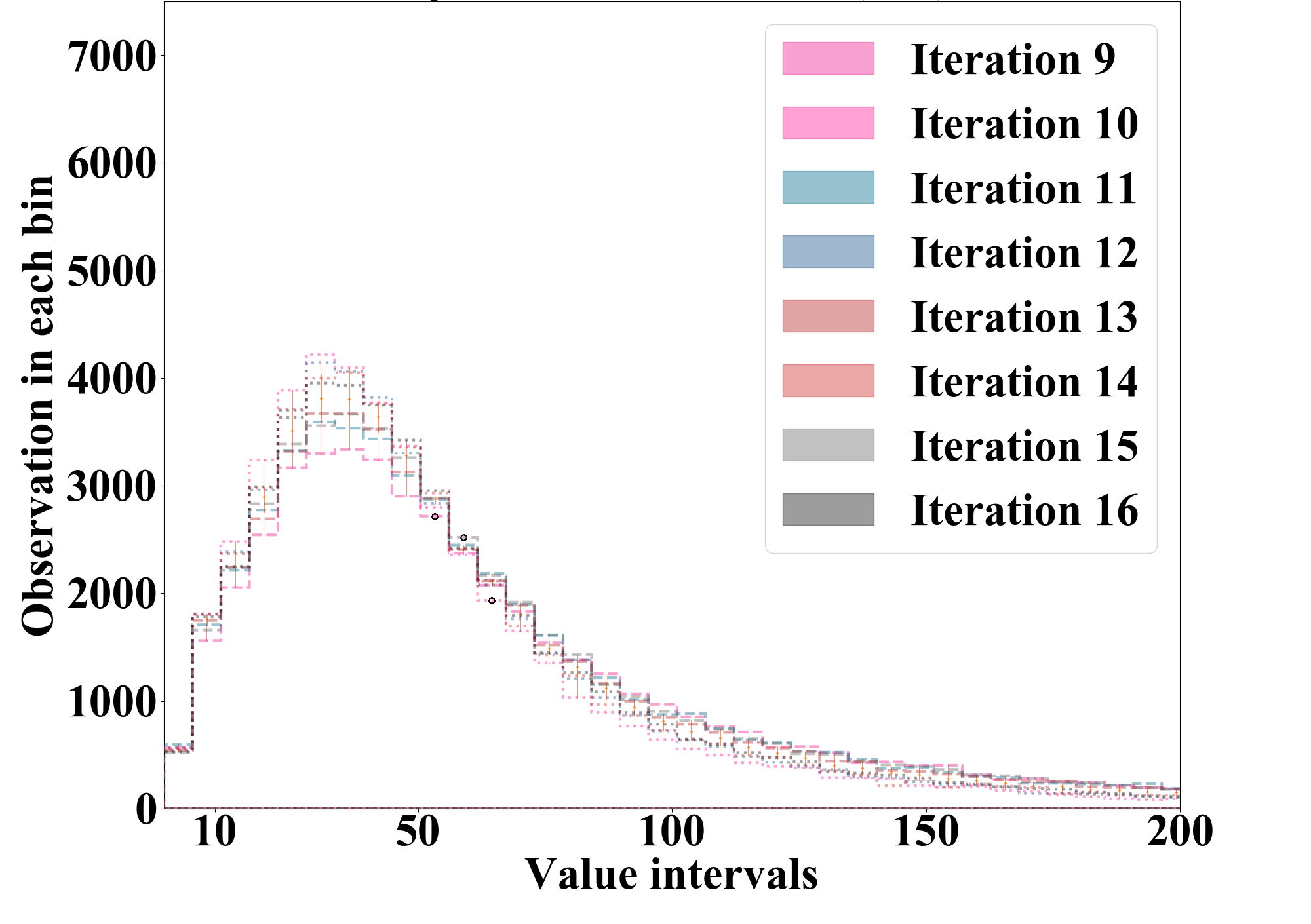}
\includegraphics[scale=0.15]{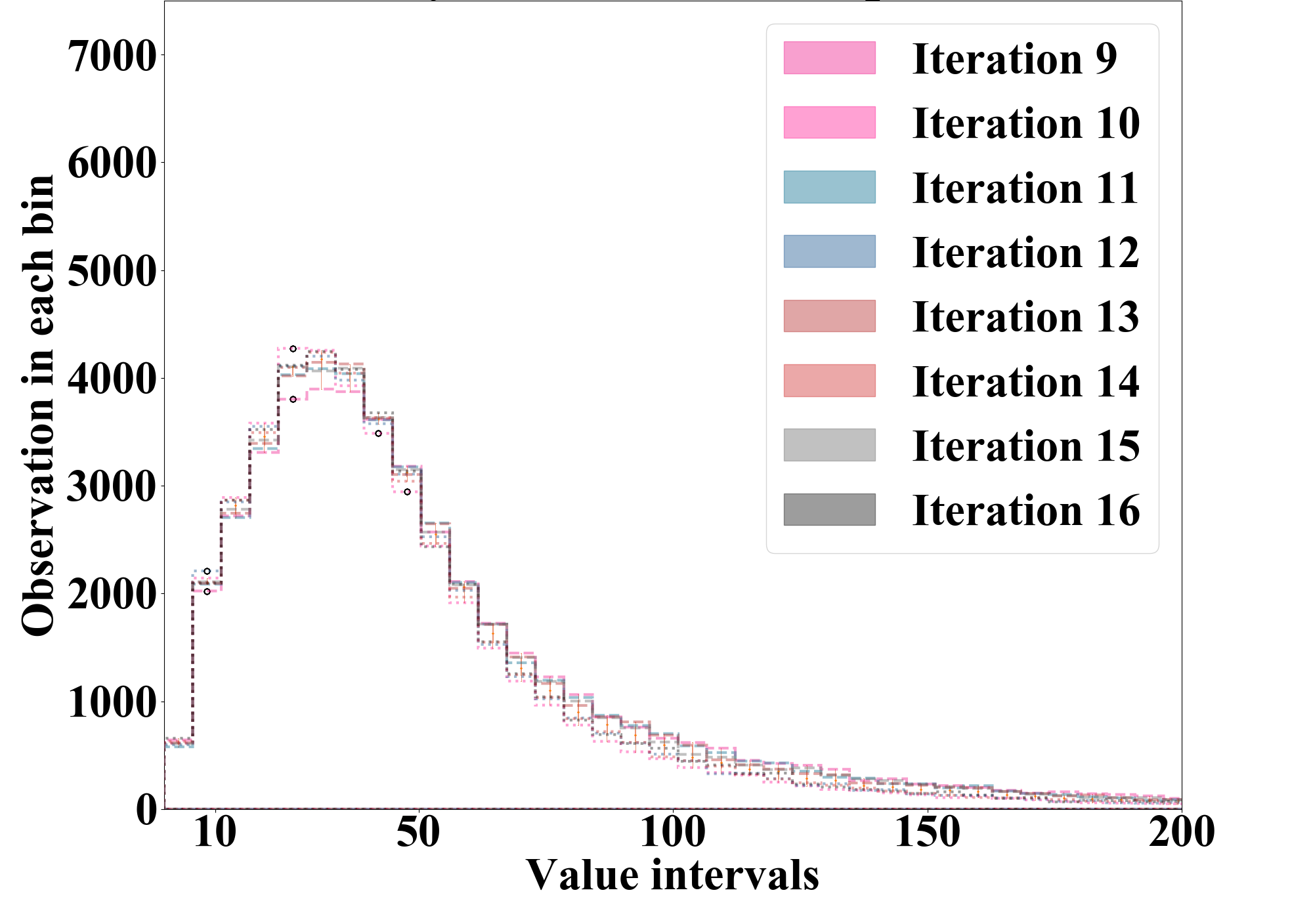}}
\captionsetup{belowskip=-10pt}
\caption{Iterations using DTA - morning (left) and afternoon peak (right)}
\label{Dynamic Iterations.png}
\end{figure}

\begin{figure}[tb]
\centerline{\includegraphics[scale=0.18]{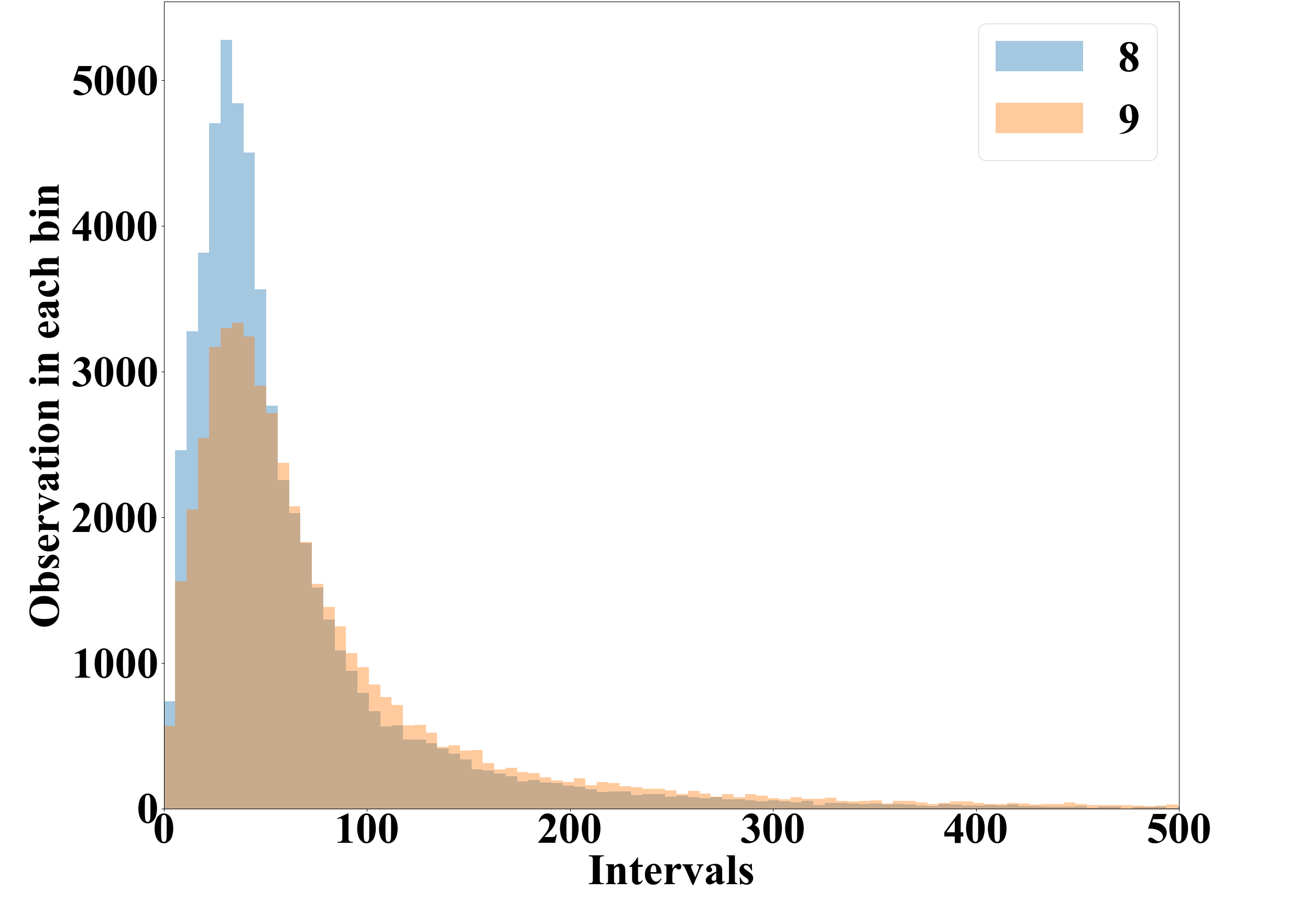}}
\captionsetup{belowskip=-10pt}
\caption{Visual representation of $\Delta$A between iterations 8 (STA) and 9 (DTA)}
\label{It 8-9.png}
\end{figure}

\begin{figure}[tb]
\centerline{\includegraphics[scale=0.13]{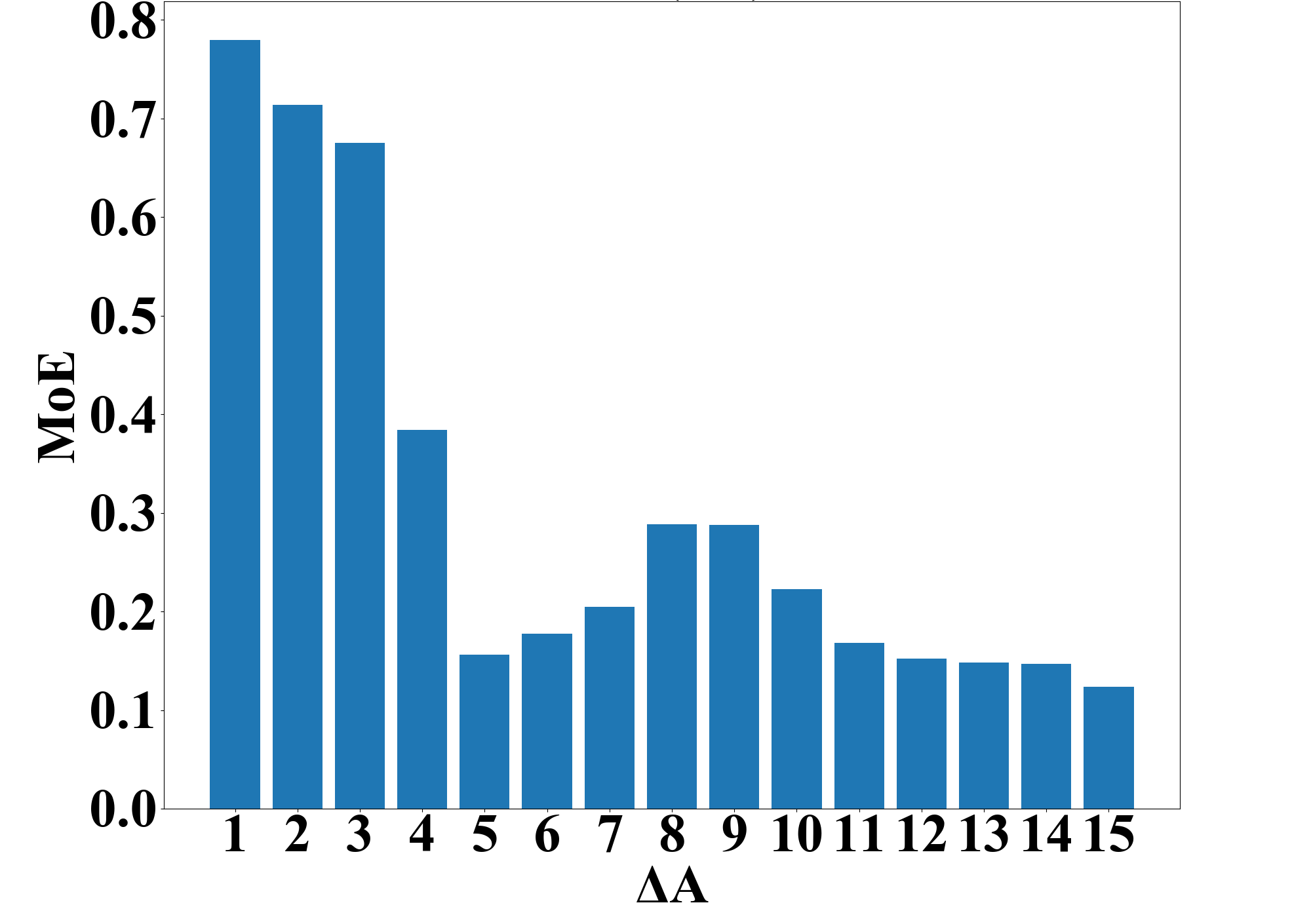}
\includegraphics[scale=0.13]{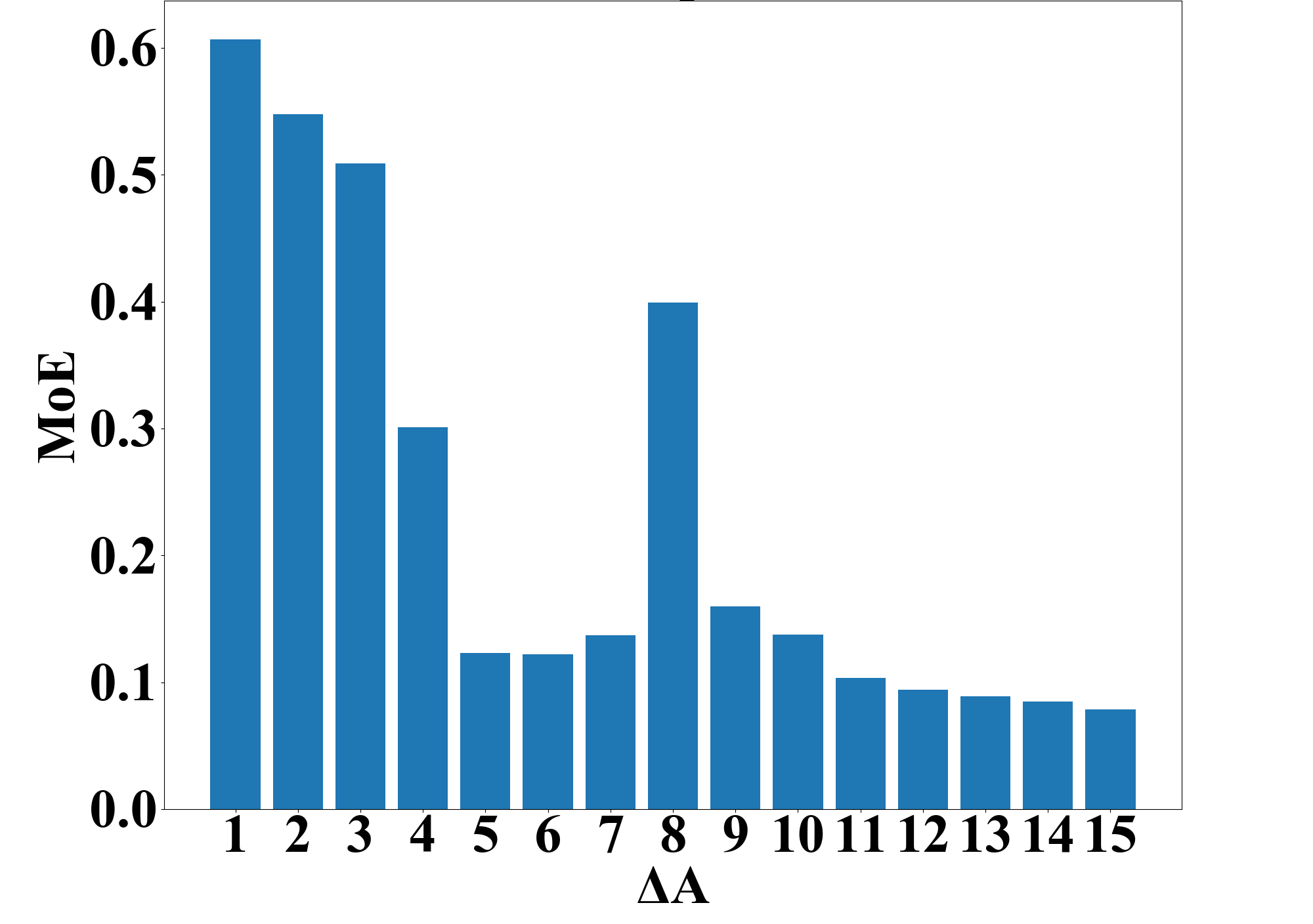}}
\captionsetup{belowskip=-10pt}
\caption{$\Delta$Tf progression - morning peak (left) and afternoon peak (right)}
\label{MoE - all.png}
\end{figure}

As it can be seen in Figure~\ref{Dynamic Iterations.png}, the stability of the iterations around the equilibrium distribution appears to be reliable, in a way that even the dynamic element of the simulations (and the resulting congestion patterns) do not disrupt said equilibrium. However, this does not imply that the DTA has no difference from the STA, as can be seen in Figure~\ref{It 8-9.png}. 
Indeed, the distribution of the $n \cdot tt$ matrix changes (as the $tt$ changes across the network due to a more realistic congestion propagation). Yet, these changes do not appear to trigger a divergent behavior (a further validation of Assumptions~\ref{2F} and~\ref{3F}). This plays a pivotal role in our method since it greatly simplifies the equilibrium search from a theoretical standpoint and implies that it is enough to perturb the skim matrix to the neighborhood of the actual equilibrium distribution, for it to then converge and remain stable in a few additional iterations.
This is reflected in the values of $\Delta$Tf for the dynamic iterations (shown in Figure~\ref{MoE - all.png}, together with the previous ones). The lowest $\Delta$Tf value for the morning is 12\% and the lowest for the afternoon is 8\%, for an average of 10\%. 

From Figure~\ref{MoE - all.png} another interesting detail emerges, as the DTA for the morning scenario has a stabilizing effect on the overall equilibrium when compared to the STA ones. This is due to the real-time adaptation of the stochastic route choice algorithm, capable of "fixing" smaller oscillations around the equilibrium in real-time (while Wardrop's first principle employed in the STA assigns vehicles on a link even if capacity is breached).

\subsection{Summary of the baseline results and validation}
While $\Delta$Tf seems able to capture the stability of the equilibrium around a $n \cdot tt$ distribution properly, none of the assumptions per se guarantee that the found equilibrium distribution is representative of the baseline. The closeness to the baseline situation (i.e., the data and distributions recorded and compared with the modeled results) is indeed more dependent on the quality of the calibration of the two single models than on the $\Delta$Tf value, perturbation, or final equilibrium distribution. To properly understand the relevance of the validation, it should be enough to focus on a case where one of the two models is not sensitive enough to one of the parameters fed back from the other model (e.g., delays do not result in the proper drop in demand). In this case, an equilibrium point would still be reached due to the assumption reported in \ref{assumptions}, but said equilibrium would frame an improper demand-supply state. 
Besides, even if the two models are calibrated properly, the validation step is still crucial to ensure that the perturbation carried out in Section~\ref{Methodology} does not steer either model too far away from the baseline, possibly to a region in Figure~\ref{Ass1} where $f_x$ has a negligible slope. The proposed MoE frames how the quantity of traffic differs between iterations, but does not provide any information about its spatio-temporal distribution. $\Delta$Tf alone cannot prove the quality of the baseline (which is still reliant on the two single models), but it is conceived instead to guarantee that the identified baseline reflects the equilibrium between demand and supply parameters. 
As it can be seen in ~\ref{Tot trips}, the total number of trips generated by the activity-based model settles very quickly around $\sim$1,120,000 legs, whereas the baseline value for the total number of trips is 1,124,810 \cite{Preprint, trb2023}, which means a negligible error. When looking only at the morning car trips, instead, the variance before iteration 5 (perturbed) is relevant (up to $\sim$40,000 legs) but it drops consistently after perturbing the $tt$ distribution in iteration 5 ($\sim$5,000 legs). Still, the total number of trips in the STA (up to iteration 8) consistently results overestimated against the baseline value of 100,094 (which results from the Aimsun OD adjustment process \cite{AimsunManual} and is based on detector counts (Figure~\ref{Aimsun Rsq.png})). This overestimation is likely caused by the mentioned inability of the STA to effectively limit the flow on links when capacity is reached and by the lack of propagation of these congestion fronts (which leads to an underestimation of travel times). But, once these aspects are taken into account with the DTA iterations, the overall number of legs in the morning settles in a neighborhood around 108,000 legs ($\pm 5,000$). This results in an error of around 7.9\%.

\begin{table}[tb]
\footnotesize
\caption{Progression of the total number of legs from the activity-based model}
\setlength\tabcolsep{1.5pt}
\begin{tabular*}{\linewidth}{@{\extracolsep\fill}lcccccccc}
\toprule%
& \multicolumn{8}{@{}c@{}}{\textbf{Iterations (STA)}}
\\\cmidrule{2-9}%
  & \textbf{\textit{1}}& \textbf{\textit{2}} & \textbf{\textit{3}}& \textbf{\textit{4}} & \textbf{\specialcell{5 \\(perturbed)}} & \textbf{\textit{6}} & \textbf{\textit{7}} & \textbf{\textit{8}}\\
\midrule
\specialcell{Total number \\ of legs}  & 1123512 & 1149755 & 1119356 & 1148166 & 1134498 & 1138192 & 1133182 & 1138958\\
\specialcell{Number of legs \\ (AM - by car)} & 92487 & 135103 & 98320 & 133274 & 118058 & 123668 & 116808 & 124911\\
\botrule
& \multicolumn{8}{@{}c@{}}{\textbf{Iterations (DTA)}}
\\\cmidrule{2-9}%
  & \textbf{\textit{9}}& \textbf{\textit{10}} & \textbf{\textit{11}}& \textbf{\textit{12}} & \textbf{\textit{13}}& \textbf{\textit{14}} & \textbf{\textit{15}} & \textbf{\textit{16}} \\
\midrule
\specialcell{Total number \\ of legs}  & 1134498 & 1111447 & 1130078 & 1114166 & 11129193 & 1114508 & 1130683 & 1116810\\
\specialcell{Number of legs \\ (AM - by car)} & 118058 & 97065 & 115155 & 101078 & 113608 & 101323 & 113801 & 103074\\
\botrule
\end{tabular*}
\label{Tot trips}
\end{table}

\begin{figure}[tb]
\centerline{\includegraphics[scale=0.40]{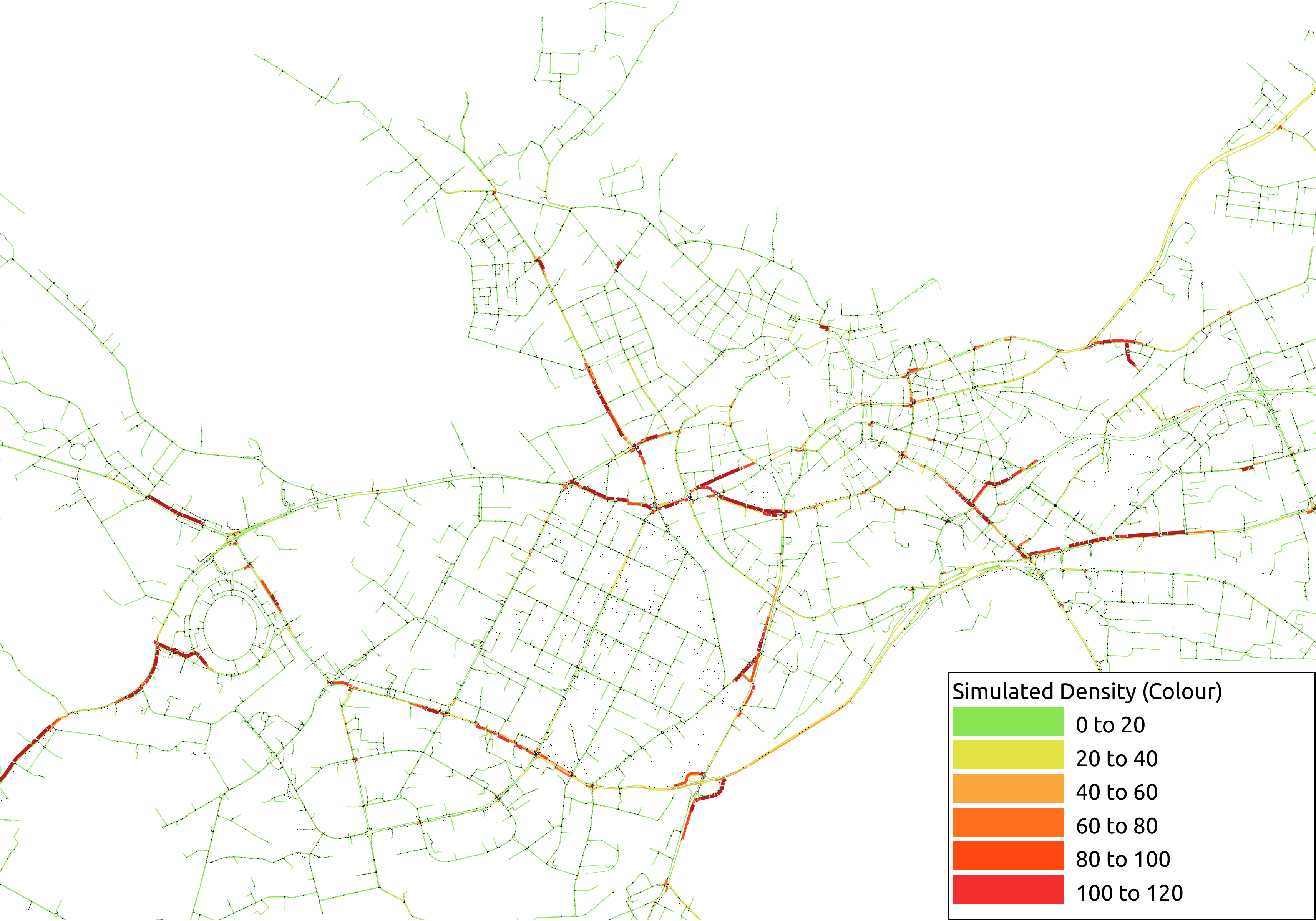}}
\captionsetup{belowskip=-10pt}
\caption{9:15 AM - Traffic density (green as free flow, red as congested)}
\label{am congestion.png}
\end{figure}

Besides, an inspection of the congestion fronts across the network at the equilibrium distribution results in expected patterns, with congestion fronts arising in the city center or on main arterial roads accessing or exiting the city (morning or afternoon peak), as in Figure~\ref{am congestion.png}.

\begin{figure}[tb]
\centerline{\includegraphics[scale=0.3]{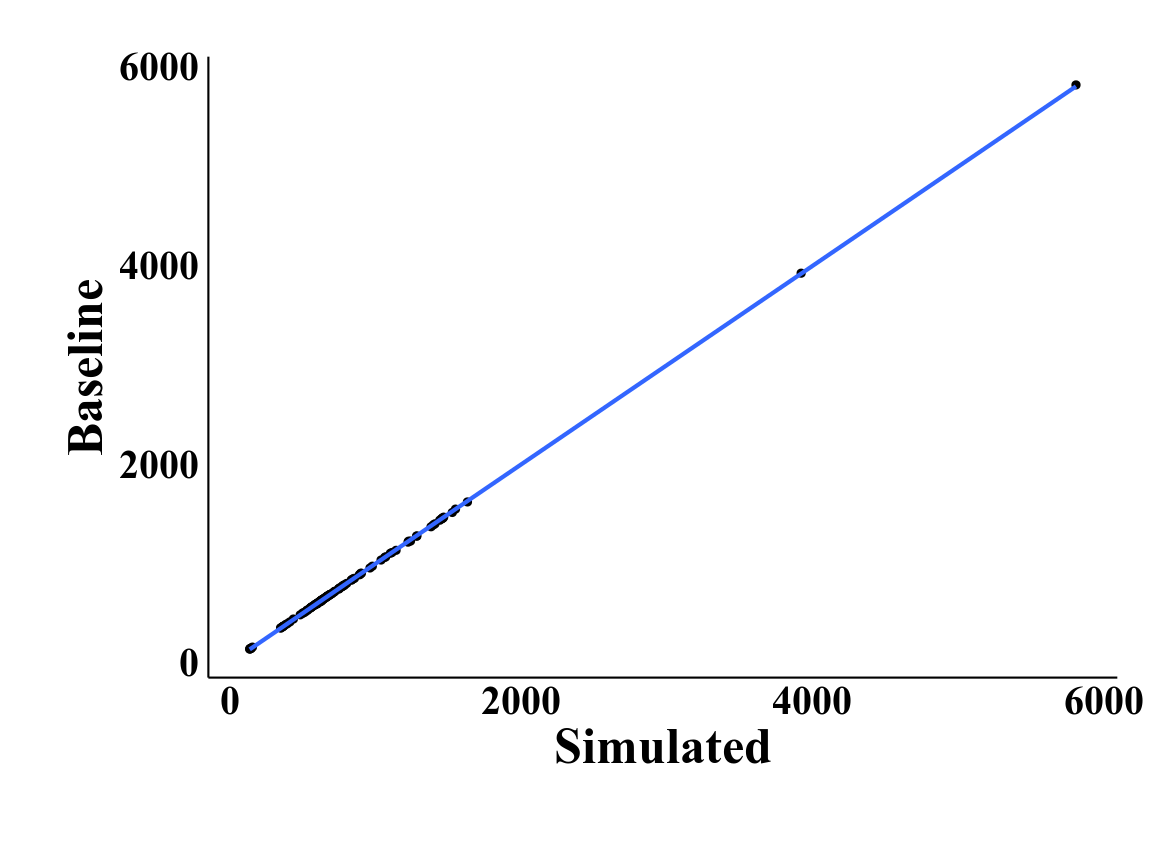}
\includegraphics[scale=0.3]{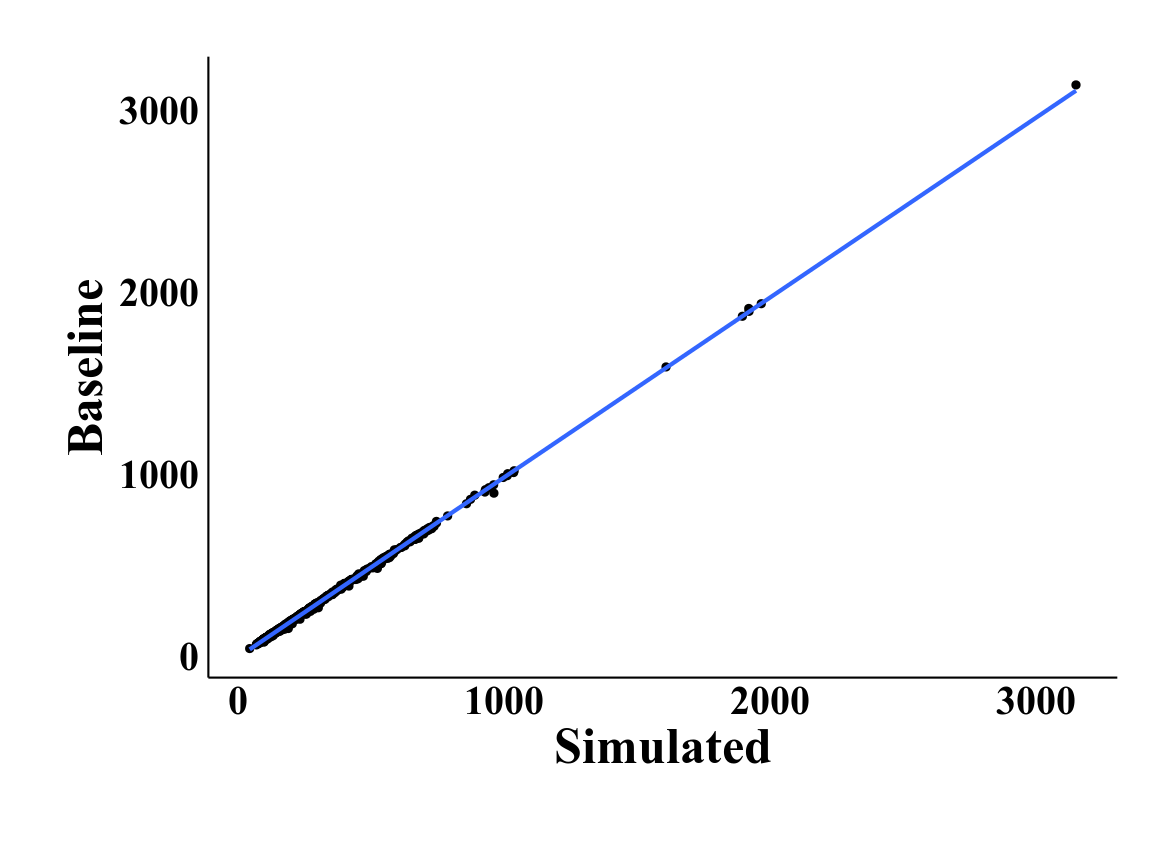}}
\centerline{\includegraphics[scale=0.30]{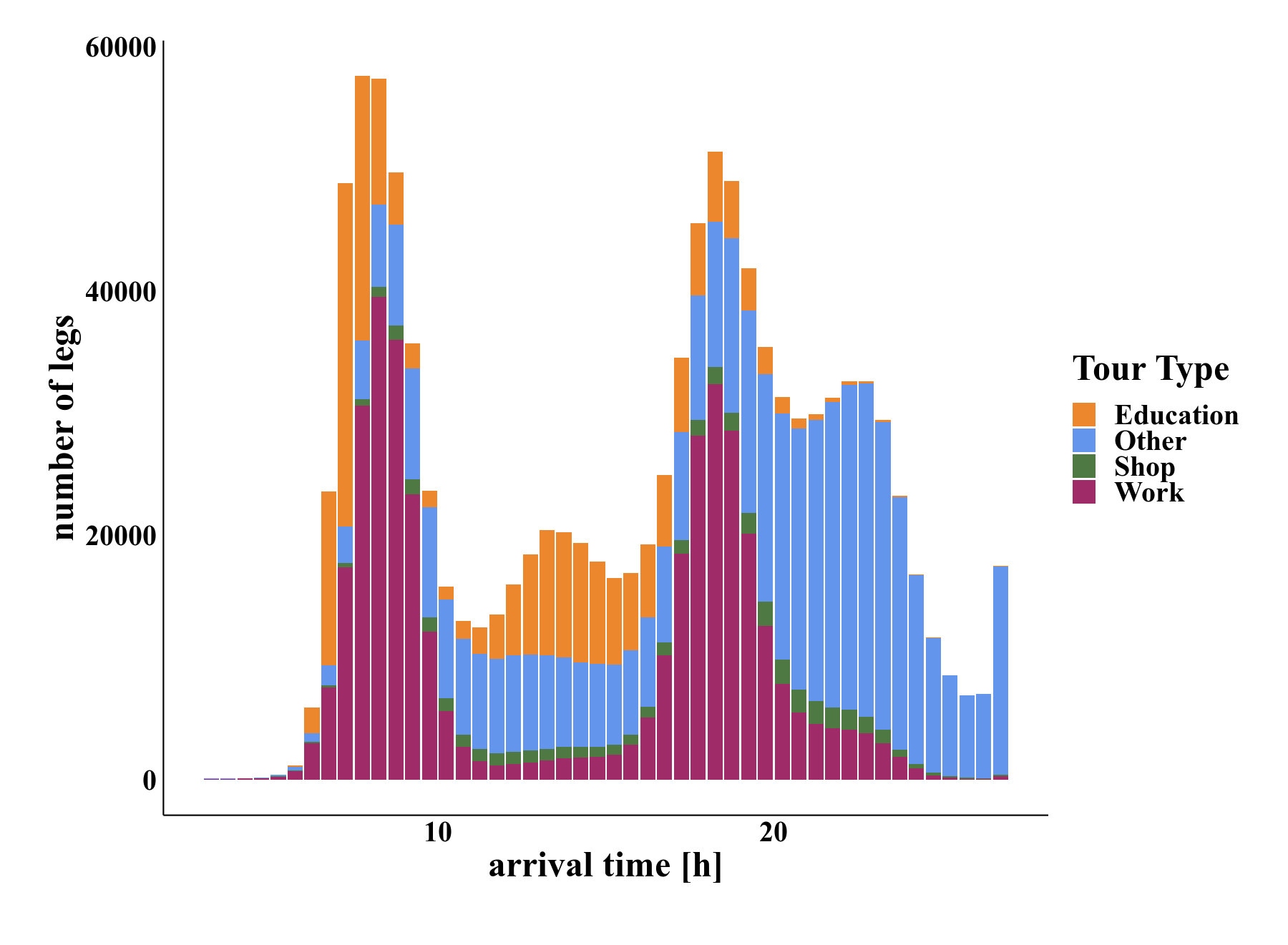}}
\captionsetup{belowskip=-10pt}
\caption{Validation of the education OD (top left), the work OD (top right), and the time distribution of the trips (bottom)}
\label{SimMobValidation}
\end{figure}

On the demand side, the daily activity schedule produced by the DTA in the last iteration is compared with ground-truth distributions such as the OD matrixes for school and work, the overall modal share, or the distribution of trips during the day. These can be found in Figure~\ref{SimMobValidation}, which shows perturbing the $tt$ does not skew the simulations away from the major mobility patterns of the baseline.

Besides, the OD at the district level arising from the simulation is compared with a baseline OD built on phone data \cite{Hadachi}. The results are judged satisfactory, considering that the measurement period in \cite{Hadachi} does not perfectly match the data with which the two models have been built (the two studies are completely unrelated).
Finally, the modal share closely matches the baseline distribution \cite{Cavoli}, as reported in Table~\ref{ModeShare}.

\begin{figure}[tb]
\centerline{\includegraphics[scale=0.14]{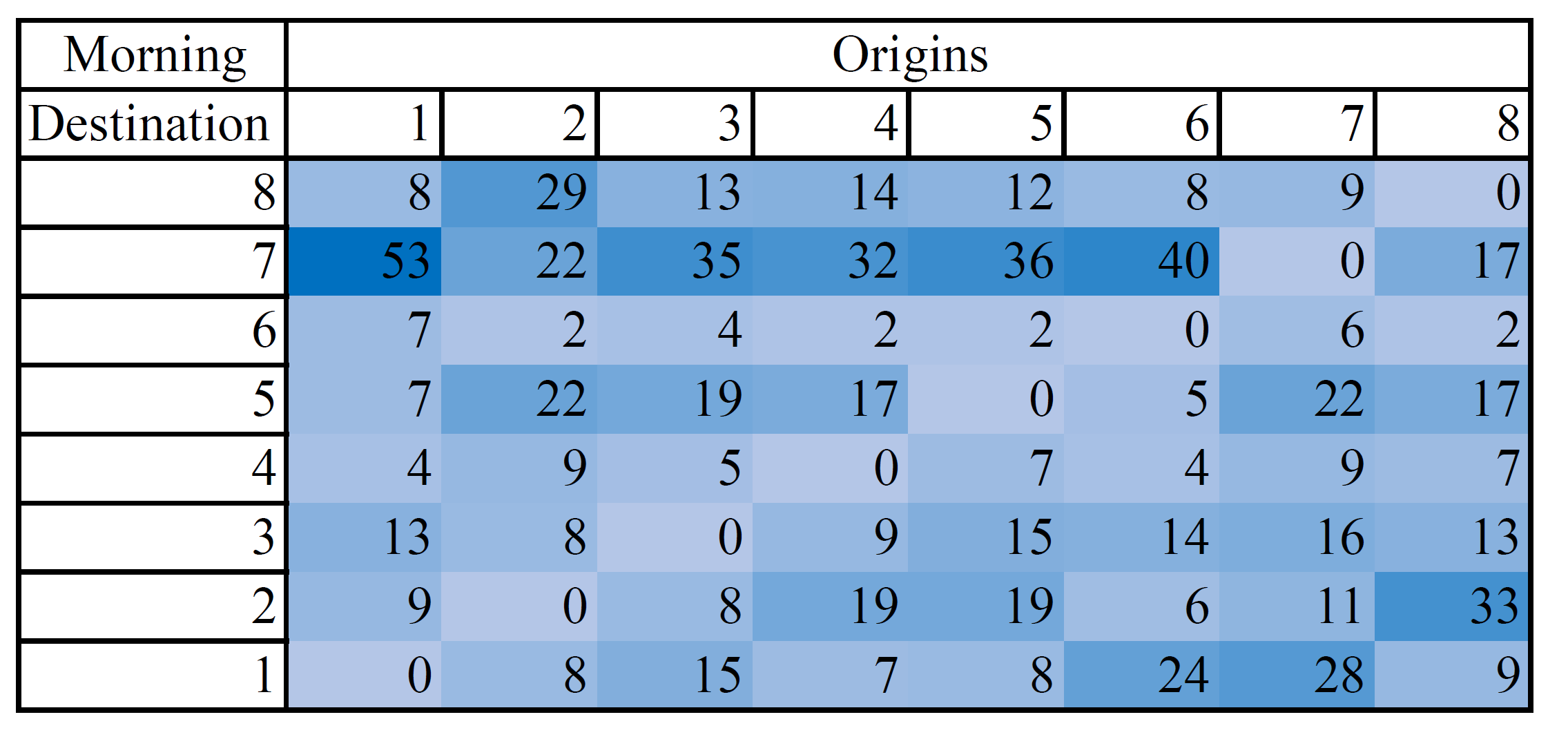}
\includegraphics[scale=0.35]{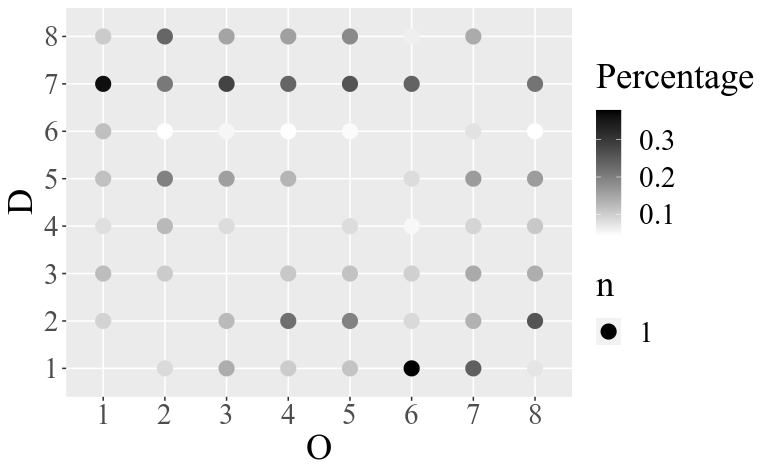}}
\centerline{\includegraphics[scale=0.14]{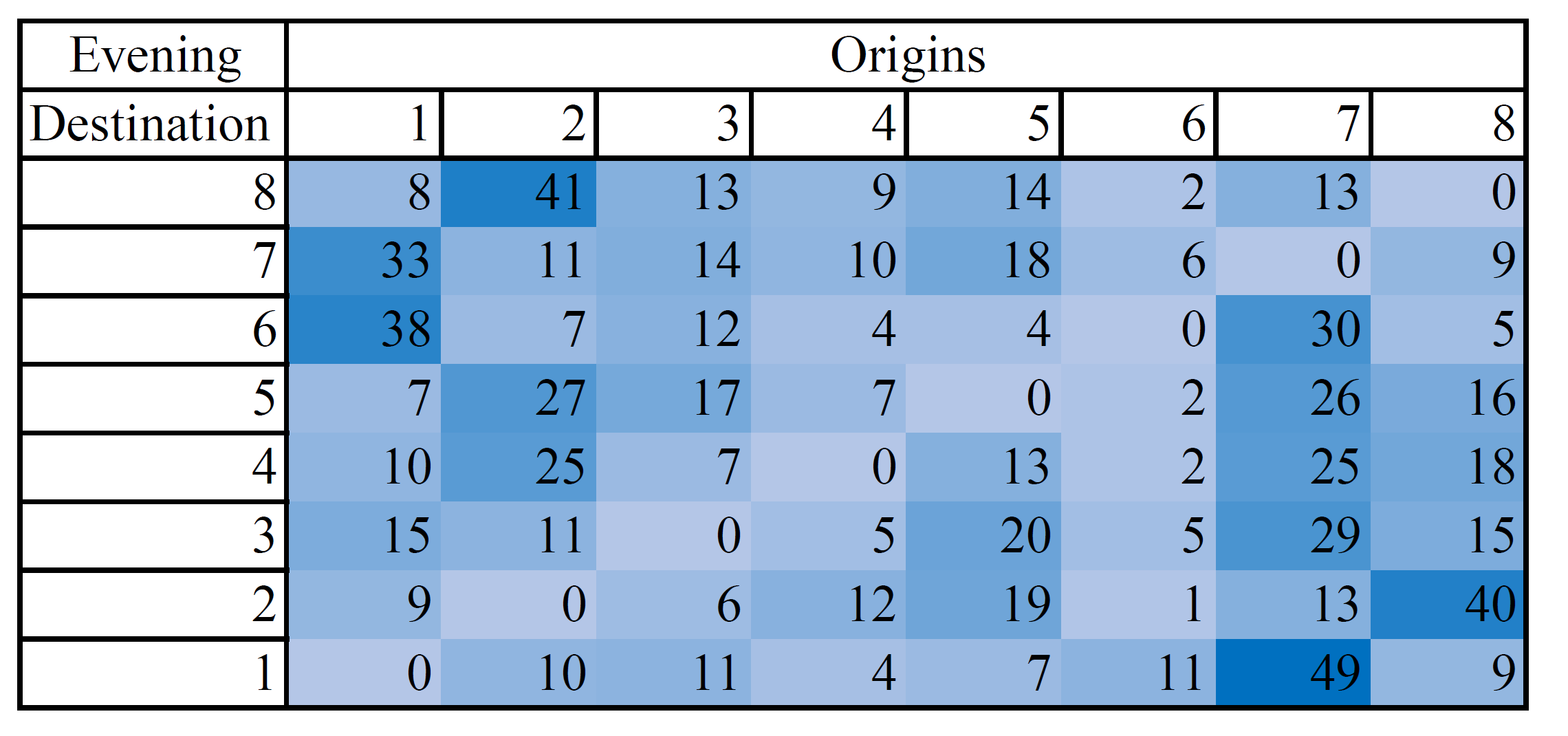}
\includegraphics[scale=0.35]{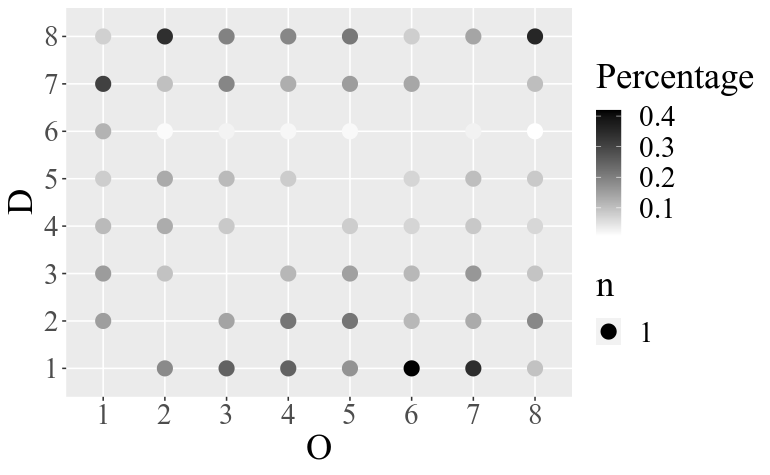}}
\captionsetup{belowskip=-10pt}
\caption{OD comparison at district level between \cite{Hadachi} (left) and simulated trips in the last DTA iteration (right) - cells are color-coded based on the \% of trips from an origin to every destination}
\label{ODdistrict}
\end{figure}

\begin{table}[tb]
\caption{Validation - Equilibrium (run 16) against the baseline \cite{Cavoli}}
\normalsize
\begin{tabular}{m{3.9cm}m{2.5cm}m{2.5cm}}
\toprule
\textbf{Parameter} & \textbf{\textit{Baseline}}& \textbf{\textit{Equilibrium}} \\
\midrule
Private vehicle share & 49\% & 49\% \\
Public transport share & 26\% & 25\% \\
Walking share & 24\% & 24\% \\
\botrule
\end{tabular}
\label{ModeShare}
\vspace{-3mm}
\end{table}

\section{Conclusions}
The paper proposed a novel approach to integrate large-scale activity-based models and traffic assignment models, with close to no requirements concerning the tools themselves. High-level requirements are defined through three assumptions, which then lead to the definition of an MoE transferable and comparable among case studies. The search space exploration is likewise detailed and the resulting equilibrium is assessed for an existing case study. Overall the paper aims to provide an easily replicable blueprint for the integration of existing models and hopefully foster wider adoption of this kind of modeling both in academia and among other stakeholders. Numerically, the case study results in negligible errors in the modal share ($\sim$1\%) and overall number of trips, $\Delta$Tf values are instead 12\% and 8\% for morning and afternoon peaks, respectively. 

\section*{Acknowledgements}
This research was funded by the FINEST Twins Center of Excellence, H2020 European Union funding for Research and Innovation grant number 856602 and the Academy of Finland project ALCOSTO (349327). 
The authors would also like to thank Lampros Yfantis and Grant Mackinnon, from Aimsun, for their valuable suggestions. 
Part of the calculations presented above were performed using computer resources within the Aalto University School of Science “Science-IT” project.

\vspace{3mm}
\bibliography{output}

\end{document}